\documentclass[12pt]{iopart}
\usepackage{graphics,iopams,mathrsfs}

\makeatletter 
\def\@fnsymbol#1{^{\thefootnote}\relax}
\makeatother

\begin{document}

\title{Superexchange Interactions in Orthorhombically
Distorted
Titanates $R$TiO$_3$ ($R$$=$ Y, Gd, Sm, and La)}

\author{I V Solovyev}

\address{Computational Materials Science Center,
National Institute for Materials Science, 1-2-1 Sengen, Tsukuba,
Ibaraki 305-0047, Japan}
\ead{solovyev.igor@nims.go.jp}

\begin{abstract}
Starting from the multiorbital Hubbard model for the $t_{2g}$-bands
of $R$TiO$_3$ ($R$$=$ Y, Gd, Sm, and La), where all the parameters
have been derived from the first-principles electronic structure
calculations, we
construct an effective superexchange (SE) spin model,
by treating transfer integrals as a perturbation. We consider
four approximations for the SE interactions:
(i) the canonical crystal-field (CF) theory,
where the form of the the occupied $t_{2g}$-orbitals is dictated by the
CF splitting at each Ti-site and three extensions of
the CF theory, namely (ii) the relativistic one,
where occupied orbitals are confined within the
lowest Kramers doublet obtained from the diagonalization of the
crystal field and relativistic spin-orbit (SO) interactions;
(iii) the finite-temperature extension, which
consider the effect of
thermal orbital fluctuations near the CF configuration on
interatomic interactions between the spins;
(iv) the many-electron extension,
which is based on the diagonalization
of the full Hamiltonian
constructed in the basis of two-electron
states separately for each bond of the system.
The main results are summarized as follows.
(i) Thermal fluctuations
of the orbital degrees of freedom
can substantially reduce the value of the magnetic transition temperature.
(ii) The relativistic SO coupling
is generally responsible for
anisotropic and antisymmetric Dzyaloshinsky-Moriya interactions.
All interactions are rigorously derived and their implications
to the magnetic properties of $R$TiO$_3$ are discussed.
(iii) The CF theory, although applicable for YTiO$_3$ and
high-temperature structures of GdTiO$_3$
and SmTiO$_3$, breaks down in the case of LaTiO$_3$.
In the latter, the CF splitting is small. Therefore, the
many-electron effects \emph{in the bonds}
as well as
the
relativistic SO interaction
start to play an important role.
It is argued that the combination of these two effects
can be responsible for the AFM
character of interatomic
correlations in LaTiO$_3$.
(iv) The SE interactions in YTiO$_3$ strongly depend
on the details of the crystal structure.
Crystal distortions
in the low-temperature structure tend to weaken the ferromagnetic interactions.
\end{abstract}

%\maketitle

\section{\label{sec:Introduction} Introduction}

  Recently, the titanium perovskites $R$TiO$_3$
(where $R$ is the rare-earth element or Y)
have attracted considerable attention.
These, formally isostructural and isoelectron materials, crystallize
in the orthorhombic space group $D^{16}_{2h}$ and have only
one electron in the $t_{2g}$-shell of Ti-atoms.
Nevertheless, the magnetic properties of
$R$TiO$_3$ appear to be extremely sensitive to the magnitude
and details of the lattice distortion.
More distorted compounds with $R$$=$ Yb - Gd
(including YTiO$_3$) appear to be
ferromagnetic (FM). The Curie temperature ($T_{\rm C}$)
typically
varies from about 30 K ($R$$=$ Y and Gd) till 60 K
($R$$=$ Dy, Ho, and Tm) \cite{Greedan,Komarek07,Knafo}.
Less distorted compounds ($R$$=$ Sm - La) form the so-called
G-type antiferromagnetic (AFM) structure, where all
nearest-neighbor (NN) spins are coupled antiferromagnetically.
The N\'eel temperature ($T_{\rm N}$) varies monotonously from 146 K
(LaTiO$_3$) till 48 K (SmTiO$_3$).
The complete phase diagram of $R$TiO$_3$ can be found in \cite{Komarek07}.

  The origin of the FM-AFM transition and,
more generally, the microscopic physical picture underlying the behavior of
these Mott-Hubbard insulators with
partially filled $t_{2g}$-shell is a matter of considerable
debates. Different types of theories ranging from the degenerate orbital
liquid \cite{KhaliullinMaekawa} to the distortion-controlled
crystal-field (CF) splitting of the
atomic
$t_{2g}$-levels \cite{MochizukiImada,PavariniNJP,Schmitz}
have been proposed.
The main obstacle with the setting of the proper physical model is
related to the fact that there are too many model parameters, which
can drastically change the picture if they are chosen in a
uncontrollable way.
For example, the orbital liquid state is rapidly deteriorated by the
lattice distortion away from the simple cubic structure.
The conventional CF theories \cite{MochizukiImada,Schmitz}
crucially depend on the choice of the dielectric constant, which
controls the magnitude of the $t_{2g}$-level splitting.
Thus, in unbiased theories for $R$TiO$_3$,
it is very important to reduce to the minimum all
arbitrariness related to the choice of the model parameters
by using
for these purposes
first-principles electronic
structure calculations.
This direction is sometimes called
``the realistic modeling of strongly correlated systems'' \cite{review2008}.

  Our previous works \cite{review2008,PRB04,PRB06b} were devoted
to construction of the parameter-free lattice fermion model
(more specifically -- the multiorbital Hubbard model)
for the low-energy $t_{2g}$-bands of distorted perovskite oxides:
\begin{equation}
\hat{\cal H}_{LF} = \sum_{ij} \sum_{\alpha \beta} h_{ij}^{\alpha \beta}
\hat{c}^\dagger_{i\alpha}\hat{c}^{\phantom{\dagger}}_{j\beta} +
\frac{1}{2}
\sum_i  \sum_{\alpha \beta \gamma \delta} U_{\alpha \beta
\gamma \delta} \hat{c}^\dagger_{i\alpha} \hat{c}^\dagger_{i\gamma}
\hat{c}^{\phantom{\dagger}}_{i\beta}
\hat{c}^{\phantom{\dagger}}_{i\delta}.
\label{eqn:Hmanybody}
\end{equation}
The model itself is specified
in the basis of six Wannier orbitals (the so-called spin-orbitals), which are denoted by
Greek symbols, each of which is a
combination of two spin
($s$$=$ $\uparrow$ or $\downarrow$) and three orbital ($m$$=$ $1$, $2$, or $3$)
variables. The site-diagonal part of $\hat{h}_{ij} \equiv \| h_{ij}^{\alpha \beta} \|$
includes the CF Hamiltonian
$\hat{h}^{\rm CF}_i$
and the relativistic spin-orbit (SO) interaction $\hat{h}^{\rm SO}_i$.
The off-diagonal ($i$$\ne$$j$) elements
of $\hat{h}_{ij}$ stand for the transfer integrals $\hat{t}_{ij} \equiv \| t_{ij}^{\alpha \beta} \|$.
Both $\hat{h}^{\rm CF}_i$ and $\hat{t}_{ij}$ are diagonal with respect
to the spin indices.
$U_{\alpha \beta \gamma \delta}$ are the matrix elements of screened
Coulomb interactions. All parameters were determined
in the \textit{ab initio} fashion, in the frameworks of the
density functional theory. Other details can be found in
the original work \cite{PRB06a}
as well as in the review article \cite{review2008}.

  In this paper we discuss how
the lattice fermion model
(\ref{eqn:Hmanybody}) for the $t_{2g}^1$ compounds can be further mapped onto the
the spin-$1/2$ model of the general form
\begin{equation}
\hat{\cal H}_S =
\sum_{\langle ij \rangle} (\hat{\bsigma}_i, \hat{A}_{ij} \hat{\bsigma}_j),
\label{eqn:spinmodel1}
\end{equation}
where $\hat{A}_{ij}$ is the three-dimensional tensor, describing various
interactions between spins at the sites
$i$ and $j$. It can be further
decomposed into isotropic Heisenberg ($J_{ij})$,
antisymmetric Dzyaloshinsky-Moriya (${\bi d}_{ij}$),
and symmetric anisotropic ($\hat{\tau}_{ij}$) interactions,\footnote{
Note that for the spin $1/2$, the single-ion anisotropy term is absent.
}
so that the model itself can be rewritten
\begin{equation}
\hat{\cal H}_S = - \sum_{\langle ij \rangle} J_{ij} (\hat{\bsigma}_i, \hat{\bsigma}_j) +
\sum_{\langle ij \rangle} ({\bi d}_{ij}, [\hat{\bsigma}_i \times \hat{\bsigma}_j]) +
\sum_{\langle ij \rangle} (\hat{\bsigma}_i, \hat{\tau}_{ij} \hat{\bsigma}_j),
\label{eqn:spinmodel2}
\end{equation}
where $\hat{\bsigma}_i = (\hat{\sigma}_i^x,\hat{\sigma}_i^y,\hat{\sigma}_i^z)$
is the vector of Pauli matrices in the spin subspace
and the prefactor $S^2$ for $S$$=$$1/2$ is included in the definition of the
model
parameters
$\hat{A}_{ij}$,
$J_{ij}$, ${\bi d}_{ij}$, and $\hat{\tau}_{ij}$.
Thus, $J_{ij}$ is the scalar,
${\bi d}_{ij} = (d_{ij}^x,d_{ij}^y,d_{ij}^z)$ is the vector, and
$\hat{\tau}_{ij} = \| \tau_{ij}^{ab} \|$
(where $a$ and $b$ stand to denote the $x$-, $y$-, or $z$-components in the orthorhombic
coordinate frame)
is the three-dimensional
tensor satisfying the condition
$\tau_{ij}^{xx} + \tau_{ij}^{yy} + \tau_{ij}^{zz} = 0$.
The notation
$(\hat{\bsigma}_i, \hat{\bsigma}_j)$ stands for the inner product
and $[\hat{\bsigma}_i \times \hat{\bsigma}_j]$ -- for the cross product
of the vectors
$\hat{\bsigma}_i$ and $\hat{\bsigma}_j$.

  The goal of this work is twofold.
On the one hand we will discuss different levels of approximations
for the superexchange (SE) interactions underlying derivation
of the spin Hamiltonian (\ref{eqn:spinmodel1}).
On the other hand, we investigate the accuracy of realistic
modeling for different types of materials. Particularly,
how far one can go with the description of not only qualitative
but also quantitative aspects of interatomic magnetic interactions
in $R$TiO$_3$.
For these purposes we will consider four
characteristic compounds: FM YTiO$_3$ and GdTiO$_3$, and
G-type AFM SmTiO$_3$ and LaTiO$_3$.
We will also present details of the
parameters
of the Hubbard Hamiltonian (\ref{eqn:Hmanybody}) and
interatomic magnetic interactions (\ref{eqn:spinmodel1}-\ref{eqn:spinmodel2}).
After brief introduction of the crystal
structure of $R$TiO$_3$ (Sec.~\ref{sec:structure}) and description of
parameters of the Hubbard Hamiltonian as derived from the first-principles
electronic structure
calculations (Sec.~\ref{sec:HubbardParameters}),
we will turn directly to the analysis of the SE interactions.
We will start with the one-electron approximation (referring to the form
of the ground-state wavefunction in the atomic limit) and consider
different types of the one-electron theories, such as the conventional
CF theory at $T$$=$$0$ (Sec.~\ref{sec:CF}), CF
theory with the relativistic SO interactions, which explains
the appearance of anisotropic and antisymmetric
Dzyaloshinsky-Moriya interactions (Sec.~\ref{sec:CFSO}). We will also
estimate the effect of thermal fluctuations of the orbital
degrees of freedom
on the interatomic interactions between the spins (Sec.~\ref{sec:MultiSlater}).
In Sec.~\ref{sec:SingleSlater} we will consider a many-electron
extension of the SE theory, where the ground-state
wavefunction
is determined from the diagonalization of two-electron
Hamiltonians constructed separately for each bond of the system.
Particularly,
we will argue that the behavior of LaTiO$_3$ is drastically differrent from other
compounds.
If in YTiO$_3$, GdTiO$_3$, and SmTiO$_3$ the type of the magnetic
ground state is mainly determined by the CF splitting, which is clearly
larger than the energy gain caused by the
virtual hoppings underlying the
SE processes
and the relativistic SO interaction,
in LaTiO$_3$ all three factors are at least comparable
making the situation less straightforward.
This revives the idea of our earlier work \cite{PRB04},\footnote{
Note however that the CF splitting calculated in \cite{PRB04}
is incorrect, because it does not take into account the nonsphericity
of the Madelung potential \cite{MochizukiImada}. Similar problem exists
in \cite{PavariniNJP}. The situation was discussed in \cite{PRB06b}.
}
and now we will explicitly show how the combination of
many-electron effects in the bonds and the relativistic SO
interaction may explain the AFM character of interatomic
correlations in LaTiO$_3$.
Finally, brief summary of the work and main conclusions
will be presented in Sec.~\ref{sec:Conclusions}.

\section{\label{sec:structure} Crystal Structure}

\begin{figure}[h!]
\centering \noindent
\resizebox{7cm}{!}{\includegraphics{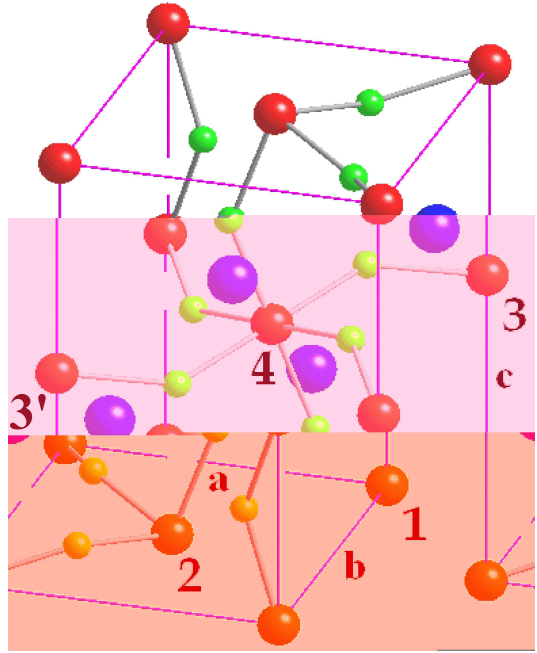}}
\caption{\label{fig.structure}
Crystal structure
of YTiO$_3$. The Y atoms are indicated
by the big blue spheres, the Ti are indicated
by the medium red spheres, and the oxygen atoms are indicated
by the small green spheres. The symbols ${\bf a}$, ${\bf b}$, and ${\bf c}$ stand for
the orthorhombic translations.
Different Ti-atoms are shown by numbers.
}
\end{figure}

  The titanates $R$TiO$_3$ ($R$$=$ Y, Gd, Sm, and La) crystallize
in the orthorhombic space group is
$D^{16}_{2h}$ in Sch\"{o}nflies notations
(No. 62 in International Tables). An example of the crystal
structure with the notations of four Ti-atoms composing the
primitive cell is shown in Fig.~\ref{fig.structure} for the case of YTiO$_3$ \cite{Maclean}.

  All calculations have been performed in the experimental crystal structure.
The low-temperature (LT) structure
is currently available only for
YTiO$_3$ and LaTiO$_3$, which was measured at $T$$=$ 2 K \cite{Komarek07} and
$T$$=$ 8 K \cite{Cwik}, respectively.
Moreover, in order to make a direct
comparison with our previous works \cite{review2008,PRB06b},
we also used another set of parameters for YTiO$_3$,
corresponding to
the room temperature (RT) \cite{Maclean}.
For SmTiO$_3$ and GdTiO$_3$,
the crystal structure was taken from \cite{Komarek07},
correspondingly for $T$$=$ 100 K and 290 K.

  Other details of the crystal structure and their implications to the
properties of $t_{2g}$-compounds can be found in the previous
publications \cite{review2008,PRB06b}.

\section{\label{sec:HubbardParameters} Parameters of Hubbard Model}

  In this section we summarize parameters of the Hubbard model (\ref{eqn:Hmanybody}),
as obtained from the first-principles electronic structure calculations.
The procedure used for the construction of the
low-energy model (\ref{eqn:Hmanybody}) as well as for the
calculation of the model parameters was described in many
details in \cite{review2008,PRB06a}.
It is true that the procedure itself relies on a number
of approximations. Nevertheless, we would like
to emphasize that apart from these approximations,
we did not use any adjustable parameters. In this sense,
the procedure is parameter-free.

  Basically, there are
three sets of the model parameters
describing
\begin{enumerate}
\item
the CF splitting of the atomic
$t_{2g}$-levels,
\item
the transfer integrals,
\item
the on-site Coulomb
and exchange interactions.
\end{enumerate}

  The behavior of parameters of the \emph{crystal field}
is explained in table~\ref{tab:CF}, which shows the eigenvalues and eigenvectors
obtained after the diagonalization of $\hat{h}_i^{\rm CF}$.
\begin{table}[h!]
\caption{Eigenenergies (measured in meV from the lowest energy)
and eigenvectors obtained after the diagonalization
of the crystal-field Hamiltonian $\hat{h}_i^{\rm CF}$.
The eigenvectors are expanded over the basis of the
$xy$, $yz$, $z^2$, $zx$, and $x^2$-$y^2$ orbitals,
in the orthorhombic coordinate frame.}
\label{tab:CF}
\begin{indented}
\item[]\begin{tabular}{lcc}
\br
  compound     & energies & orbitals         \\
\mr
 & $0$ &
$-0.13,   \phantom{-}0.45,   \phantom{-}0.38,   -0.61,   \phantom{-}0.51$
\\
YTiO$_3$ (RT)  & $109$ &
$\phantom{-}0.46,    \phantom{-}0.70,    \phantom{-}0.14,    \phantom{-}0.08,   -0.52$
\\
 & $123$ &
$\phantom{-}0.06,   -0.46,   \phantom{-}0.32,   -0.61,  -0.56$
\\
\mr
 & $0$ &
$-0.19,   \phantom{-}0.40,   \phantom{-}0.33,   -0.57,   \phantom{-}0.61$
\\
YTiO$_3$ (LT)  & $144$ &
$\phantom{-}0.10,   -0.33,   \phantom{-}0.37,   -0.66,   -0.57$
\\
 & $226$ &
$\phantom{-}0.43,   \phantom{-}0.80,   \phantom{-}0.15,   \phantom{-}0.09,   -0.39$
\\
\mr
 & $0$ &
$-0.17,   \phantom{-}0.43,   \phantom{-}0.29,   -0.55,   \phantom{-}0.64$
\\
GdTiO$_3$ & $126$ &
$\phantom{-}0.17,   -0.14,   \phantom{-}0.37,   -0.67,   -0.60$
\\
 & $191$ &
$\phantom{-}0.38,   \phantom{-}0.85,   \phantom{-}0.04,   \phantom{-}0.21,   -0.30$
\\
\mr
 & $0$ &
$-0.16,   \phantom{-}0.50,   \phantom{-}0.21,   -0.42,   \phantom{-}0.71$
\\
SmTiO$_3$ & $112$ &
$\phantom{-}0.13,   -0.09,   \phantom{-}0.38,   -0.77,   -0.49$
\\
 & $167$ &
$\phantom{-}0.38,   \phantom{-}0.82,   \phantom{-}0.02,   \phantom{-}0.21,   -0.37$
\\
\mr
 & $0$ &
$-0.06,   -0.85,   -0.15,   \phantom{-}0.34,   -0.37$
\\
LaTiO$_3$ & $37$ &
$\phantom{-}0.20,   \phantom{-}0.46,   -0.26,   \phantom{-}0.78,   -0.28$
\\
 & $61$ &
$-0.27,   -0.15,   -0.17,   \phantom{-}0.41,   \phantom{-}0.84$
\\
\br
\end{tabular}
\end{indented}
\end{table}
One can clearly see that the CF splitting
tends to quench the orbital degrees
of freedom for all considered compounds, except LaTiO$_3$:
the lowest level, split off by the crystal field, is nondegenerate
and separated from the next level by an energy gap of
at least
100 meV. Thus, in the atomic limit, the single
electron occupies the lowest $t_{2g}$-level, and
the degeneracy of the ground state is lifted by the crystal field.
The CF splitting decreases in the direction
YTiO$_3$$\rightarrow$GdTiO$_3$$\rightarrow$SmTiO$_3$$\rightarrow$LaTiO$_3$.
Especially, in LaTiO$_3$ the CF splitting is exceptionally small.
In YTiO$_3$, there is a clear dependence of
the CF splitting on the crystal
structure. Basically, there are two
effects: (1) the order of the middle and highest $t_{2g}$-levels is
reversed in the RT structure (in comparison with the LT one);
(2) the CF splitting itself is
substantially smaller in the RT structure. The structure of
$t_{2g}$-orbitals, which are split by the crystal field, is rather
similar in YTiO$_3$, GdTiO$_3$, and SmTiO$_3$: apart from some
numerical differences,
the coefficients
of the expansion over the basis function have similar values and similar phases.
The same tendencies are clearly seen in
the RT structure of YTiO$_3$ after
the interchanging of the middle and highest
$t_{2g}$-levels. Nevertheless, the structure of the $t_{2g}$-levels in
LaTiO$_3$ appears to be different.

  The behavior of \emph{transfer integrals} in the NN
bonds is explained in table~\ref{tab:transferCF}.
\begin{table}[h!]
\caption{Transfer integrals in the bonds 1-2 and 1-3 (measured in meV)
in the local coordinate frame corresponding to the diagonal
representation of the crystal-field
splitting.
The positions of the atomic sites are explained in Fig.~\ref{fig.structure}.
Transfer integrals in other bonds can be obtained
from $\hat{t}_{12}$ and $\hat{t}_{13}$ using
the symmetry operations of the space group $D^{16}_{2h}$.}
\label{tab:transferCF}
\begin{indented}
\item[]\begin{tabular}{lcc}
\br
  compound     & $\hat{t}_{12}$ & $\hat{t}_{13}$         \\
\mr
YTiO$_3$ (RT)  &
$
\left(
\begin{array}{rrr}
  -8  &  91 &  13 \\
  -4  & -66 &  0  \\
  193 &  11 & -32 \\
\end{array}
\right)
$
&
$
\left(
\begin{array}{rrr}
 -29 & 71 & -12   \\
  71 & 82 & -38   \\
 -12 & -38 & 102  \\
\end{array}
\right)
$
\\
\mr
YTiO$_3$ (LT)  &
$
\left(
\begin{array}{rrr}
 -36  & 26 &  96  \\
  185 & -8 &  39  \\
 -26  & -7 & -68  \\
\end{array}
\right)
$
&
$
\left(
\begin{array}{rrr}
 -49 & -12 &  64  \\
 -12 &  97 & -33  \\
  64 & -33 & 114  \\
\end{array}
\right)
$
\\
\mr
GdTiO$_3$ &
$
\left(
\begin{array}{rrr}
 -48  &  54 &  85 \\
  171 &  6  &  23 \\
 -56  & -20 & -79 \\
\end{array}
\right)
$
&
$
\left(
\begin{array}{rrr}
 -34 &  13 &  80  \\
  13 &  98 & -24  \\
  80 & -24 & 143  \\
\end{array}
\right)
$
\\
\mr
SmTiO$_3$ &
$
\left(
\begin{array}{rrr}
  -83 &  43 &  81  \\
  159 &  39 &  27  \\
  -45 & -23 & -84  \\
\end{array}
\right)
$
&
$
\left(
\begin{array}{rrr}
 -10  &  1  & 101  \\
  1   & 114 & -12  \\
  101 & -12 & 143  \\
\end{array}
\right)
$
\\
\mr
LaTiO$_3$ &
$
\left(
\begin{array}{rrr}
 -62 & 130 &  94  \\
 -14 & -5  &  94  \\
  76 & 76  & -69  \\
\end{array}
\right)
$
&
$
\left(
\begin{array}{rrr}
  142 & -70 & 78 \\
 -70  &  90 & 53 \\
  78  &  53 & 53 \\
\end{array}
\right)
$
\\
\br
\end{tabular}
\end{indented}
\end{table}
The transfer integrals are presented in the local coordinate frame of
orbitals, which diagonalize the crystal field (see table~\ref{tab:CF}).

  The \emph{screened on-site Coulomb interactions} for the $t_{2g}$-band
are expressed through the 3$\times$3$\times$3$\times$3
matrices, in the basis of three Wannier orbitals~\cite{review2008,PRB06b,PRB06a}.
For the distorted compounds,
the structure of these matrices is rather complex
and may involve many independent parameters.
These matrices are directly
used in the next section for the analysis of SE interactions
without any additional approximations or simplifications.
Nevertheless, just for explanatory purposes in this section,
the full matrix $\| U_{\alpha \beta \gamma \delta} \|$
was fitted
in terms of two
Kanamori parameters \cite{Kanamori}: the Coulomb interaction ${\cal U}$ between the
same $t_{2g}$-orbitals and the exchange interaction ${\cal J}$.
The fitting implies that
$\| U_{\alpha \beta \gamma \delta} \|$
has the same symmetry as in the spherical environment
of isolated atoms. For example, in the
process of fitting, it was assumed that the Coulomb interaction ${\cal U}'$
between different $t_{2g}$-orbitals is related to ${\cal U}$ and ${\cal J}$
by the identity ${\cal U}' = {\cal U} -2{\cal J}$ \cite{Kanamori}, and
the parameter
${\cal U}$ is the same
for all three $t_{2g}$-orbitals.
The results of such fitting are shown in table~\ref{tab:Kanamori}.
\begin{table}[h!]
\caption{Results of fitting of the
effective Coulomb interactions
in terms of two Kanamori parameters: the intraorbital Coulomb interaction ${\cal U}$
and the exchange interaction ${\cal J}$.
All energies are measured in eV.}
\label{tab:Kanamori}
\begin{indented}
\item[]\begin{tabular}{@{}lccc}
\br
 compound      & ${\cal U}$    & ${\cal J}$ \\
\mr
 YTiO$_3$ (RT) &    $3.45$     & $0.62$     \\
 YTiO$_3$ (LT) &    $3.40$     & $0.62$     \\
 GdTiO$_3$     &    $3.28$     & $0.62$     \\
 SmTiO$_3$     &    $3.29$     & $0.62$     \\
 LaTiO$_3$     &    $3.20$     & $0.61$     \\
\br
\end{tabular}
\end{indented}
\end{table}
The screening of the Coulomb interaction
${\cal U}$ is sensitive to the local environment in solids:
generally, ${\cal U}$ is larger for the more distorted YTiO$_3$ and smaller for the least
distorted LaTiO$_3$~\cite{review2008}. On the contrary, the
exchange interaction ${\cal J}$ is less
sensitive to the details of the screening and
close to the atomic limit~\cite{PRL05}.

\section{\label{sec:SingleSlater}Superexchange Interactions in the One-Electron
Approximation}

  The superexchange interaction in the bond $i$-$j$ is
related to
the gain of the kinetic energy, which is acquired by an
electron residing at the atomic site $i$ in the
process of virtual hoppings in the subspace of unoccupied
orbitals at the atomic site $j$, and vice versa \cite{PWA,KugelKhomskii}.
Let us consider first the approximation adopted in \cite{PRB06b,KO2},
where
the energy gain caused by virtual hoppings
in the bond $i$-$j$
was computed in the following way:
\begin{equation}
\fl
{\cal T}(\varphi_i,\varphi_j) = - \left\langle G
\left| \hat{t}_{ij} \left( \sum_M \frac{{\hat{\mathscr{P}}}_j|j M
\rangle \langle j M|
{\hat{\mathscr{P}}}_j}{E_{jM}} \right) \hat{t}_{ji} +
(i \leftrightarrow j) \right| G \right\rangle.
\label{eqn:egain}
\end{equation}
Namely, we start with the lattice of isolated atoms, each of which accommodates
one electron, and describe the ground-state wavefunction of such a reference system
by the single Slater determinant $G$.
Since in the atomic limit there is only one $t_{2g}$-electron per one Ti-site,
this is essentially one-electron problem and all many-electron effects emerge
only in the process of virtual hoppings. Typically, this justifies the use of the
single-determinant approximation for $G$ in
the case of titanates \cite{PRB06b}.
By denoting the occupied one-electron
orbitals at the sites $i$ and $j$ as
$\varphi_i$ and $\varphi_j$, respectively,
the Slater determinant $G$ for the bond $i$-$j$
takes the following form:
$$
|G(1,2) \rangle = \frac{1}{\sqrt{2}} \left[ \varphi_i(1) \varphi_j(2) -
\varphi_i(2) \varphi_j(1) \right],
$$
where the symbols ``1'' and ``2'' stand for the coordinates of two electrons,
associated with the sites $i$ and $j$.
After that we treat the transfer integrals $\hat{t}_{ij}$
as a perturbation.
Since the Coulomb repulsion ${\cal U}$ (and ${\cal U}'$) is large, it is
sufficient to consider the excited configurations accommodating only two
$t_{2g}$-electrons, which contribute to the second order perturbation theory
with respect to $\hat{t}_{ij}$. The eigenvalues and eigenvectors
of such excited configurations at the site $j$ are denotes as
$E_{jM}$ and $|jM \rangle$, respectively.
They
are
obtained after the diagonalization of the Coulomb interactions,
$$
\frac{1}{2}\sum_{\alpha \beta \gamma \delta} U_{\alpha \beta \gamma \delta}
\hat{c}^\dagger_{i\alpha} \hat{c}^\dagger_{i\gamma}
\hat{c}^{\phantom{\dagger}}_{i\beta}
\hat{c}^{\phantom{\dagger}}_{i\delta},
$$
in the basis
of all possible two-electron Slater determinants at the site $j$.
For six $t_{2g}$ spin-orbitals, there are
$6(6$$-$$1)/2$$=$$15$ such determinants \cite{PRB06b}.
Thus, each $|jM \rangle$ is constructed from several Slater
determinants and takes into account the correct multiplet
structure in the atomic limit.
Examples of the atomic multiplet structures corresponding to the
exciting two-electron configurations can be
found in \cite{PRB06b}.
$\hat{\mathscr{P}}_j$ in (\ref{eqn:egain}) is a projector
operator, which enforces the Pauli principle and suppresses any
transfer of an electron into
the subspace of occupied orbitals at the site $j$.
In practice, $\hat{\mathscr{P}}_j$ projects $|jM \rangle$
into the subspace spanned by the Slater determinants of the form
$$
|G_{i \to j}(1,2) \rangle = \frac{1}{\sqrt{2}}
\left[ \psi_j(1) \varphi_j(2) -
\psi_j(2) \varphi_j(1) \right],
$$
where $\varphi_j$ is the occupied orbital at the site $j$
and $\psi_j$ is any orbital residing at the site $j$, which
can be reached from $\varphi_i$ by the transfer integrals $\hat{t}_{ij}$.

  Thus, the expression (\ref{eqn:egain}) for the energy gain
combines elements of the
one-electron approximation for $|G \rangle$ with the exact many-electron treatment
for the excited two-electron states.
On the one hand,
since $|jM \rangle$ is a
combination of several Slater determinants, the method takes into account
some many-electron effects in the atomic limit, which may have interesting consequences
on the magnetic properties of $R$TiO$_3$.\footnote{
For example, in the case of the AFM spin alignment in the bond
$i$-$j$, the excited configuration
$\uparrow$$\downarrow$ is decomposed into
two-electron singlet and triplet states, that leads to the
additional energy gain in equation (\ref{eqn:egain}). This effect
additionally stabilizes the AFM interactions, as is manifested
for example
in somewhat stronger canting of spin magnetic moments
away from the collinear FM alignment
in the ground state of YTiO$_3$ in comparison with results of the
mean-field Hartree-Fock approximation \cite{review2008,PRB06b}.
}
On the other hand, the form of
$|G \rangle$ is restricted by the single Slater determinant.
In this sense, this is an one-electron approach.
That is why, in the following we will refer to the expression (\ref{eqn:egain})
as to an one-electron approximation for the SE interactions. The
multi-determinant analog of (\ref{eqn:egain}) will be
considered in Sec. \ref{sec:MultiSlater}.

\subsection{\label{sec:CF}Crystal-Field Theory for the
Superexchange Interactions at $T$$=$$0$}

  In the crystal-field theory, it is assumed that the form of the
occupied orbitals $\{ \varphi_i \}$ is totally controlled by the
crystal field, which is much larger than the energy gain caused
by the virtual hoppings (\ref{eqn:egain}) as well as the energies
of thermal fluctuations. Then,
in the
one-electron approximation,
the isotropic spin
couplings $J_{ij}$
is related to the energy gains (\ref{eqn:egain})
by the following expression
$$
2 J_{ij} =
{\cal T}^{\uparrow \uparrow}_{ij} -
{\cal T}^{\uparrow \downarrow}_{ij},
$$
where
${\cal T}^{\uparrow \uparrow}_{ij} \equiv {\cal T}(\varphi^\uparrow_i,\varphi^\uparrow_j)$
and
${\cal T}^{\uparrow \downarrow}_{ij} \equiv {\cal T}(\varphi^\uparrow_i,\varphi^\downarrow_j)$, and
the indices $\uparrow$ and $\downarrow$ explicitly show the
spin states of the occupied orbitals $\varphi_i$ and $\varphi_j$.
The values of $J_{ij}$ between the nearest and some next nearest
neighbors
are listed in table \ref{tab:CFJ}.
It also shows the corresponding type of the magnetic ground state expected in the
$R$TiO$_3$ compounds.
\begin{table}[h!]
\caption{Results of the crystal-field theory for $R$TiO$_3$:
isotropic superexchange interactions ($J_{ij}$, measured in meV),
corresponding magnetic transition temperatures ($T_{\rm C,N}$, measured in K) obtained in the
random phase and mean-field approximation (results of the mean-field
approximation are shown parenthesis), and the type of the
magnetic the ground state (GS, where ``F'' stands for the ferromagnetic state
and ``A'' - for the A-type antiferromagnetic state).
The positions of the atomic sites are explained in Fig.~\ref{fig.structure}.
Depending on the magnetic ground state, the notations $T_{\rm C}$ and $T_{\rm N}$
stand for the Curie and N\'eel temperature, respectively.}
\label{tab:CFJ}
\begin{indented}
\item[]\begin{tabular}{lcccccc}
\br
  compound     & $J_{12}$ & $J_{13}$          &  $J_{23}$ & $J_{23'}$ & $T_{\rm C,N}$ &    GS             \\
\mr
YTiO$_3$ (RT)  &   $3.23$ & $\phantom{-}0.45$ &  $0.03$   & $-0.16$   & 74 (154)    &     F              \\
YTiO$_3$ (LT)  &   $2.93$ &           $-0.14$ &  $0.03$   & $-0.16$   & 85 (144)    &     A              \\
GdTiO$_3$      &   $2.79$ & $\phantom{-}0.68$ &  $0.07$   & $-0.17$   & 79 (138)    &     F              \\
SmTiO$_3$      &   $0.68$ & $\phantom{-}1.60$ &  $0.07$   & $-0.03$   & 41 ~(68)    &     F              \\
LaTiO$_3$      &   $1.41$ &           $-4.88$ &  $0.30$   & $\phantom{-}0.13$   & 93 (164)    &     A              \\
\br
\end{tabular}
\end{indented}
\end{table}
The magnetic transition temperature is estimated in
the random phase approximation (RPA, \ref{sec:appendixA}).

  Thus, in the CF theory, the ground state appears to be FM
in the case of GdTiO$_3$ and SmTiO$_3$, and A-type AFM
in the case of LaTiO$_3$.
The ground state of YTiO$_3$ depends on the crystal structure:
FM and A-type AFM for the RT and LT structure, respectively.
Nevertheless, as we will see in the next section,
both structures yield the same type of the noncollinear
magnetic ground state, which combines elements of the FM and A-type AFM
ordering within one magnetic structure, when the relativistic SO
interaction is taken into account.
To certain extent the FM interactions in $R$TiO$_3$
can be also stabilized by considering explicitly the $e_g$-band
of these compounds \cite{MochizukiImada}. Nevertheless, at the present stage
the situation is not completely clear because
the same effects
would act against the G-type AFM ground state in
the case of SmTiO$_3$ and LaTiO$_3$.

  The magnetic interactions in YTiO$_3$ (RT) are in reasonable agreement
with the ones obtained in the Hartree-Fock approximation for the
Hubbard model (\ref{eqn:Hmanybody}) with the
same parameters of the crystal structure
($J_{12}$$=$ $3.2 \div 3.9$ meV and $J_{13}$$=$ $1.0 \div 1.2$ meV, depending on the
magnetic state \cite{PRB06b}).
To certain extend, the same is true for LaTiO$_3$, although
since the CF splitting is small and allows for the additional change of the orbital ordering
in each magnetic state \cite{KugelKhomskii},
the parameters of interatomic magnetic interactions
obtained in the Hartree-Fock approximation for LaTiO$_3$
exhibit a strong dependence on the magnetic state in which they
are calculated ($J_{12}$$=$ $1.0 \div 4.5$ meV and $J_{13}$$=$ $-$$1.2 \div -$$4.9$ meV \cite{PRB06b}).
Nevertheless, results of the CF theory (table \ref{tab:CFJ})
are within the parameters range obtained
previously in the Hartree-Fock approximation for LaTiO$_3$ \cite{PRB06b}.

  RPA considerably reduces the magnetic transition temperature
in comparison with the mean-field approach.
This effect is particularly strong in YTiO$_3$ due to the
quasi-two-dimensional character of the SE interactions,
which according to the Mermin-Wagner theorem suppress
the long-range magnetic order at finite $T$~\cite{MerminWagner}.\footnote{
Strictly speaking, this behavior does not seem to be consistent
with the experimental inelastic neutron scattering data, which are
typically interpreted in terms of the three-dimensional
isotropic Heisenberg model \cite{Ulrich2002}.
However, at present there is no clear consensus on this matter,
neither on theoretical nor on experimental side.
For example, the experimental orbital ordering pattern
determined in \cite{Akimitsu} is more consistent
with the anisotropic (quasi-two-dimensional) structure
of interatomic magnetic interactions \cite{PRB06b,SawadaTerakura}.
Thus, the problem requires additional study, both on
theoretical and experimental sides.
}
Nevertheless, $J_{13}$ is finite and the reduction of
the magnetic transition temperature has only logarithmic
dependence of the anisotropy $g= |J_{13}/J_{12}|$
of NN interactions: $T_{\rm C,N} \sim -$$(\ln g)^{-1}$~\cite{Nagaev}.\footnote{
Moreover, the dependence $T_{\rm C,N} \sim -$$(\ln g)^{-1}$ can be further
modified by the longer range interactions $J_{23}$ and $J_{23'}$
between neighboring ${\bf ab}$-planes.
}
Therefore, the transition temperature remains finite.
Moreover, the Curie temperature is overestimated
by factor two even in the case of YTiO$_3$.
The correct G-type AFM ground state
is reproduced neither in SmTiO$_3$ nor in LaTiO$_3$.

\subsection{\label{sec:CFSO}Relativistic Spin-Orbit Interaction and
the Effective Spin Hamiltonian}

  In the relativistic generalization of the crystal-field theory, the
occupied orbital $ \varphi_i $ at each Ti-site
is obtained after the diagonalization
of the one-electron Hamiltonian $\hat{h}^{\rm CF}_i + \hat{h}^{\rm SO}_i$,
which combines the crystal field and the
relativistic SO interaction
$\hat{h}^{\rm SO}_i = (\xi_i/2) (\bsigma_i,{\bf l}_i)$, where $\xi_i \approx 20$ meV
is related to the spherical part of the one-electron potential \cite{review2008}.
The Hamiltonian is constructed in the basis of six
$t_{2g}$ spin-orbitals. The lowest eigenstate
obtained after the diagonalization is the Kramers doublet, whose eigenvectors
can be formally denotes as $\varphi_i^1$ and $\varphi_i^2$. Then,
we consider a liner combination
$$
\varphi_i^{\pm a} = c_i^1 \varphi_i^1 + c_i^2 \varphi_i^2,
$$
and find the coefficients $c_i^1$ and $c_i^2$ from the condition
that
the averaged spin moment,
${\bi e}_i^{\pm a}$$=$$\langle \varphi_i^{\pm a} | \hat{\bsigma}_i | \varphi_i^{\pm a} \rangle$
corresponding to $\varphi_i^{\pm a}$, has the
maximal projection along the
$\pm a =$ $\pm$$x$, $\pm$$y$, and $\pm$$z$
axes in the orthorhombic coordinate frame.\footnote{
The actual procedure was based on the numerical maximization of the
function
$({\bi e}_i^{\pm a},{\bi e}_0^{\pm a})/({\bi e}_i^{\pm a},{\bi e}_i^{\pm a})^{1/2}$,
where
${\bi e}_0^{\pm x} = (\pm 1,0,0)$, ${\bi e}_0^{\pm y} = (0,\pm 1,0)$, and
${\bi e}_0^{\pm z} = (0,0,\pm 1)$.
The procedure does not uniquely specify the phase of $\varphi_i^{\pm a}$.
Nevertheless, the tensor $\hat{A}_{ij}$
of interatomic magnetic interactions does not
depend on this phase.
}
Then, we use these orbitals in the expression (\ref{eqn:egain})
for the energy gain, and calculate 36 parameters
${\cal T}(\varphi_i^{\pm a},\varphi_j^{\pm b})$ corresponding to
all possible combinations of $\pm$$a$ and $\pm$$b$ at the
sites $i$ and $j$. In the one-electron (mean-field) approximation,
the magnetic part of ${\cal T}(\varphi_i^{\pm a},\varphi_j^{\pm b})$
should correspond to the energies, which are obtained from
(\ref{eqn:spinmodel1}) by replacing the Pauli matrices $\hat{\bsigma}_i$
by the unit vectors
${\bi e}_i^\pm$, describing
different directions of spin at the site $i$.
In total,
the magnetic part of ${\cal T}(\varphi_i^{\pm a},\varphi_j^{\pm b})$
is characterized by 9 independent parameters, which
constitute the tensor $\hat{A}_{ij}$ of interatomic magnetic
interactions. The values of $\hat{A}_{ij}$ for two
inequivalent NN bonds are given in table \ref{tab:Atensor}
and the decomposition of $\hat{A}_{ij}$
into $J_{ij}$,
${\bi d}_{ij}$,
and $\hat{\tau}_{ij}$ is presented in table \ref{tab:Adecomposition}.
\begin{table}[h!]
\caption{Tensors of magnetic interactions associated with the
nearest-neighbor bonds 1-2 and 1-3 (measured in meV) as
obtained in the crystal-field theory with the relativistic
spin-orbit interaction.
The positions of the atomic sites are explained in Fig.~\ref{fig.structure}.
The magnetic interactions in other nearest-neighbor bonds can be obtained
from $\hat{A}_{12}$ and $\hat{A}_{13}$ using
the symmetry operations of the space group $D^{16}_{2h}$.}
\label{tab:Atensor}
\begin{indented}
\item[]\begin{tabular}{lcc}
\br
  compound     & $\hat{A}_{12}$ & $\hat{A}_{13}$         \\
\mr
YTiO$_3$ (RT)  &
$
\left(
\begin{array}{ccc}
      -3.11  & \phantom{-}0.22  & \phantom{-}0.06  \\
      \phantom{-}0.22  &  -3.11  & \phantom{-}0.30  \\
      \phantom{-}0.07  &  -0.12  &  -3.26  \\
\end{array}
\right)
$
&
$
\left(
\begin{array}{ccc}
      -0.38  & -0.07  & -0.25 \\
      -0.07  & -0.39  & -0.37 \\
      \phantom{-}0.25  & \phantom{-}0.37  & -0.46 \\
\end{array}
\right)
$
\\
\mr
YTiO$_3$ (LT)  &
$
\left(
\begin{array}{ccc}
      -2.89  & \phantom{-}0.26  & -0.33  \\
      -0.01  & -2.86  & \phantom{-}0.48  \\
      \phantom{-}0.41  & -0.37  & -2.97  \\
\end{array}
\right)
$
&
$
\left(
\begin{array}{ccc}
      \phantom{-}0.19  & -0.01  & -0.10 \\
      -0.01  & \phantom{-}0.18  & -0.31 \\
      \phantom{-}0.10  & \phantom{-}0.31  & \phantom{-}0.17 \\
\end{array}
\right)
$
\\
\mr
GdTiO$_3$   &
$
\left(
\begin{array}{ccc}
      -2.79  & \phantom{-}0.40  & -0.59 \\
      -0.09  & -2.66  & \phantom{-}0.62 \\
      \phantom{-}0.70  & -0.42  & -2.84 \\
\end{array}
\right)
$
&
$
\left(
\begin{array}{ccc}
      -0.62  & -0.03  & -0.22 \\
      -0.03  & -0.62  & -0.34 \\
      \phantom{-}0.22  & \phantom{-}0.34  & -0.65 \\
\end{array}
\right)
$
\\
\mr
SmTiO$_3$   &
$
\left(
\begin{array}{ccc}
      -0.66  & \phantom{-}0.72  & -1.07 \\
      -0.41  & -0.56  & \phantom{-}0.93 \\
      \phantom{-}1.22  & -0.72  & -0.72 \\
\end{array}
\right)
$
&
$
\left(
\begin{array}{ccc}
      -1.49  & -0.05  & -0.13 \\
      -0.05  & -1.57  & -0.34 \\
      \phantom{-}0.13  & \phantom{-}0.34  & -1.58 \\
\end{array}
\right)
$
\\
\mr
LaTiO$_3$   &
$
\left(
\begin{array}{ccc}
      -0.86  & -1.22  & -1.24 \\
      \phantom{-}1.84  & -1.20  &  \phantom{-}0.82 \\
       \phantom{-}0.24  & -1.71  & -0.93 \\
\end{array}
\right)
$
&
$
\left(
\begin{array}{ccc}
      \phantom{-}4.81  & \phantom{-}0.34  & -0.56 \\
      \phantom{-}0.34  & \phantom{-}4.21  & \phantom{-}3.36 \\
      \phantom{-}0.56  & -3.36  & \phantom{-}4.43 \\
\end{array}
\right)
$
\\
\br
\end{tabular}
\end{indented}
\end{table}
\begin{table}[h!]
\caption{Isotropic Heisenberg interactions ($J_{ij}$), antisymmetric
Dzyaloshinsky-Moriya interactions (${\bi d}_{ij}$), and symmetric
anisotropic ($\hat{\tau}_{ij}$) interactions associated with the nearest-neighbor
bonds 1-2 and 1-3 as
obtained in the crystal-field theory with the relativistic
spin-orbit interaction.
All interactions are measured in meV.
The positions of the atomic sites are explained in Fig.~\ref{fig.structure}.}
\label{tab:Adecomposition}
\begin{indented}
\item[]\begin{tabular}{lccc}
\br
  compound     & $J_{12}$  & ${\bi d}_{12}$ & $\hat{\tau}_{12}$         \\
\mr
YTiO$_3$ (RT)  &
$\phantom{-}3.16$
&
$
\left(
\begin{array}{c}
-0.21 \\  -0.01 \\  0 \\
\end{array}
\right)
$
&
$
\left(
\begin{array}{ccc}
      \phantom{-}0.05 &  \phantom{-}0.22 &  \phantom{-}0.07 \\
      \phantom{-}0.22 &  \phantom{-}0.05 &  \phantom{-}0.10 \\
      \phantom{-}0.07 &  \phantom{-}0.10 &   -0.10          \\
\end{array}
\right)
$
\\
\mr
YTiO$_3$ (LT)  &
$\phantom{-}2.90$
&
$
\left(
\begin{array}{c}
-0.42 \\  -0.37 \\ -0.13 \\
\end{array}
\right)
$
&
$
\left(
\begin{array}{ccc}
      \phantom{-}0.02  & \phantom{-}0.12  & \phantom{-}0.04 \\
      \phantom{-}0.12  & \phantom{-}0.05  & \phantom{-}0.06 \\
      \phantom{-}0.04  & \phantom{-}0.06  & -0.07           \\
\end{array}
\right)
$
\\
\mr
GdTiO$_3$   &
$\phantom{-}2.76$
&
$
\left(
\begin{array}{c}
-0.52 \\  -0.65 \\ -0.25 \\
\end{array}
\right)
$
&
$
\left(
\begin{array}{ccc}
      -0.02  & \phantom{-}0.15  & \phantom{-}0.05 \\
      \phantom{-}0.15  & \phantom{-}0.10  & \phantom{-}0.10 \\
      \phantom{-}0.05  & \phantom{-}0.10  & -0.08 \\
\end{array}
\right)
$
\\
\mr
SmTiO$_3$   &
$\phantom{-}0.65$
&
$
\left(
\begin{array}{c}
-0.83 \\  -1.15 \\ -0.56 \\
\end{array}
\right)
$
&
$
\left(
\begin{array}{ccc}
      -0.02  & \phantom{-}0.16  & \phantom{-}0.07 \\
      \phantom{-}0.16  & \phantom{-}0.09  & \phantom{-}0.10 \\
      \phantom{-}0.07  & \phantom{-}0.10  & -0.07 \\
\end{array}
\right)
$
\\
\mr
LaTiO$_3$   &
$\phantom{-}0.99$
&
$
\left(
\begin{array}{c}
      -1.26  \\ -0.74  \\ \phantom{-}1.53 \\
\end{array}
\right)
$
&
$
\left(
\begin{array}{ccc}
      \phantom{-}0.14  & \phantom{-}0.31 &  -0.50 \\
      \phantom{-}0.31  & -0.20  & -0.45 \\
      -0.50 &  -0.45  & \phantom{-}0.06 \\
\end{array}
\right)
$
\\
\br
\end{tabular}
\end{indented}
\begin{indented}
\item[]\begin{tabular}{lccc}
\br
  compound     & $J_{13}$  & ${\bi d}_{13}$ & $\hat{\tau}_{13}$         \\
\mr
YTiO$_3$ (RT)  &
$\phantom{-}0.41$
&
$
\left(
\begin{array}{c}
 \phantom{-}0.37  \\ -0.25  \\  0 \\
\end{array}
\right)
$
&
$
\left(
\begin{array}{ccc}
      -0.03 &  -0.07  &  0 \\
      -0.07 &  -0.02  &  0 \\
       0  &  0  & \phantom{-}0.05 \\
\end{array}
\right)
$
\\
\mr
YTiO$_3$ (LT)  &
$-0.18$
&
$
\left(
\begin{array}{c}
 \phantom{-}0.31  \\ -0.10  \\  0 \\
\end{array}
\right)
$
&
$
\left(
\begin{array}{ccc}
      \phantom{-}0.01  & -0.01  &  0 \\
      -0.01  & 0  &  0 \\
       0  &  0  & -0.01 \\
\end{array}
\right)
$
\\
\mr
GdTiO$_3$   &
$\phantom{-}0.63$
&
$
\left(
\begin{array}{c}
 \phantom{-}0.34  \\ -0.22  \\  0 \\
\end{array}
\right)
$
&
$
\left(
\begin{array}{ccc}
      \phantom{-}0.01 &  -0.03 &   0 \\
      -0.03 &  \phantom{-}0.01 &   0 \\
       0  &  0  & -0.02 \\
\end{array}
\right)
$
\\
\mr
SmTiO$_3$   &
$\phantom{-}1.55$
&
$
\left(
\begin{array}{c}
 \phantom{-}0.34  \\ -0.13  \\  0 \\
\end{array}
\right)
$
&
$
\left(
\begin{array}{ccc}
      \phantom{-}0.06  & -0.05  &  0 \\
      -0.05  & -0.03  &  0 \\
       0  &  0  & -0.03 \\
\end{array}
\right)
$
\\
\mr
LaTiO$_3$   &
$-4.48$
&
$
\left(
\begin{array}{c}
 -3.36  \\ -0.56  \\  0 \\
\end{array}
\right)
$
&
$
\left(
\begin{array}{ccc}
      -0.33 &  -0.34  &  0 \\
      -0.34 &  \phantom{-}0.27  &  0 \\
       0  &  0  & \phantom{-}0.06 \\
\end{array}
\right)
$
\\
\br
\end{tabular}
\end{indented}
\end{table}
One can clearly see that the magnetic interactions
${\bi d}_{ij}$ and $\hat{\tau}_{ij}$ of the relativistic
origin steadily increase in the direction
YTiO$_3$$\rightarrow$GdTiO$_3$$\rightarrow$SmTiO$_3$$\rightarrow$LaTiO$_3$,
which is quite consistent with the decrease of the CF splitting
in the same direction (see table~\ref{tab:CF}).
Nevertheless, it is interesting to note that the relativistic
SO interaction
also contributes to the isotropic
interactions $J_{ij}$,
and
all of them decrease (with some exception of LaTiO$_3$)
after taking into account the SO interaction
in comparison with results of the pure CF theory in table~\ref{tab:CFJ}.

  The directions of the magnetic moments in the
ground state and the
values of the
magnetic transition temperature in the
mean-field approximation (\ref{sec:appendixB}) are summarized in table \ref{tab:SOGS}.
\begin{table}[h!]
\caption{Magnetic ground state, direction of magnetization
in the ground state
and the magnetic transition temperature ($T_{\rm C,N}$, measured in K)
in the mean-field approximation as obtained in the
crystal-field theory with the relativistic spin-orbit interaction.
The vector of magnetization is referred to the site 1
in Fig.~\protect\ref{fig.structure}. Similar vectors at other Ti-sites
of the primitive cell are obtained by applying the symmetry
operations of the space group $D^{16}_{2h}$.}
\label{tab:SOGS}
\begin{indented}
\item[]\begin{tabular}{lccc}
\br
  compound     & ground state & direction of magnetization &  $T_{\rm C,N}$ \\
\mr
YTiO$_3$ (RT)  & G-A-F & $(-0.02,-0.54,\phantom{-}0.84)$ & 164    \\
YTiO$_3$ (LT)  & G-A-F & $(-0.05,-0.89,\phantom{-}0.45)$ & 148    \\
GdTiO$_3$      & G-A-F & $(-0.12,-0.38,\phantom{-}0.92)$ & 150    \\
SmTiO$_3$      & G-A-F & $(-0.35,-0.20,\phantom{-}0.91)$ & \phantom{1}94    \\
LaTiO$_3$      & C-F-A & $(-0.21,-0.47,\phantom{-}0.86)$ & 197    \\
\br
\end{tabular}
\end{indented}
\end{table}
As expected, the magnetic ground state is noncollinear.
The type of the
magnetic ground state
for the space group $D_{2h}^{16}$
can be formally denoted as
X-Y-Z, where X, Y, and Z is the magnetic structure
(F, A, C, or G) formed by the projections of the magnetic moments
onto the orthorhombic axes ${\bf a}$, ${\bf b}$, and ${\bf c}$,
respectively. By comparing the values of the magnetic
transition temperature with the ones reported in table \ref{tab:CFJ},
one can clearly see that the SO interaction tends to additionally increase
$T_{\rm C}$, despite the fact that all isotropic interactions
$J_{ij}$ decrease. Moreover, the SO interaction alone
does not solve the problem of the magnetic ground state of LaTiO$_3$,
which is expected to be of the C-F-A type and does not includes the
experimentally observed G-component. The magnetic ground state of
other compounds is of the G-A-F type. Particularly, in YTiO$_3$,
there is a considerable weight of the both A- and F-components
parallel to the orthorhombic ${\bf b}$- and ${\bf c}$-axes, respectively. The weight
of the G-component parallel to the ${\bf a}$-axis is negligibly small.
The main difference between the RT and LT structures is in the
relative weight of the A- and F-components (the formed is larger
in the LT structure). On the other hand, in SmTiO$_3$ there is a
substantial weight of the G-type AFM component along the ${\bf a}$-axis.
The magnetic structures obtained for YTiO$_3$ (RT)
and LaTiO$_3$ are consistent
results of the Hartree-Fock calculations, which were considered
in the previous publication \cite{PRB06b}.

\subsection{\label{sec:T}Thermal Fluctuations of the Orbital
Degrees of Freedoms}

\begin{figure}[h!]
\centering \noindent
\resizebox{6.5cm}{!}{\includegraphics{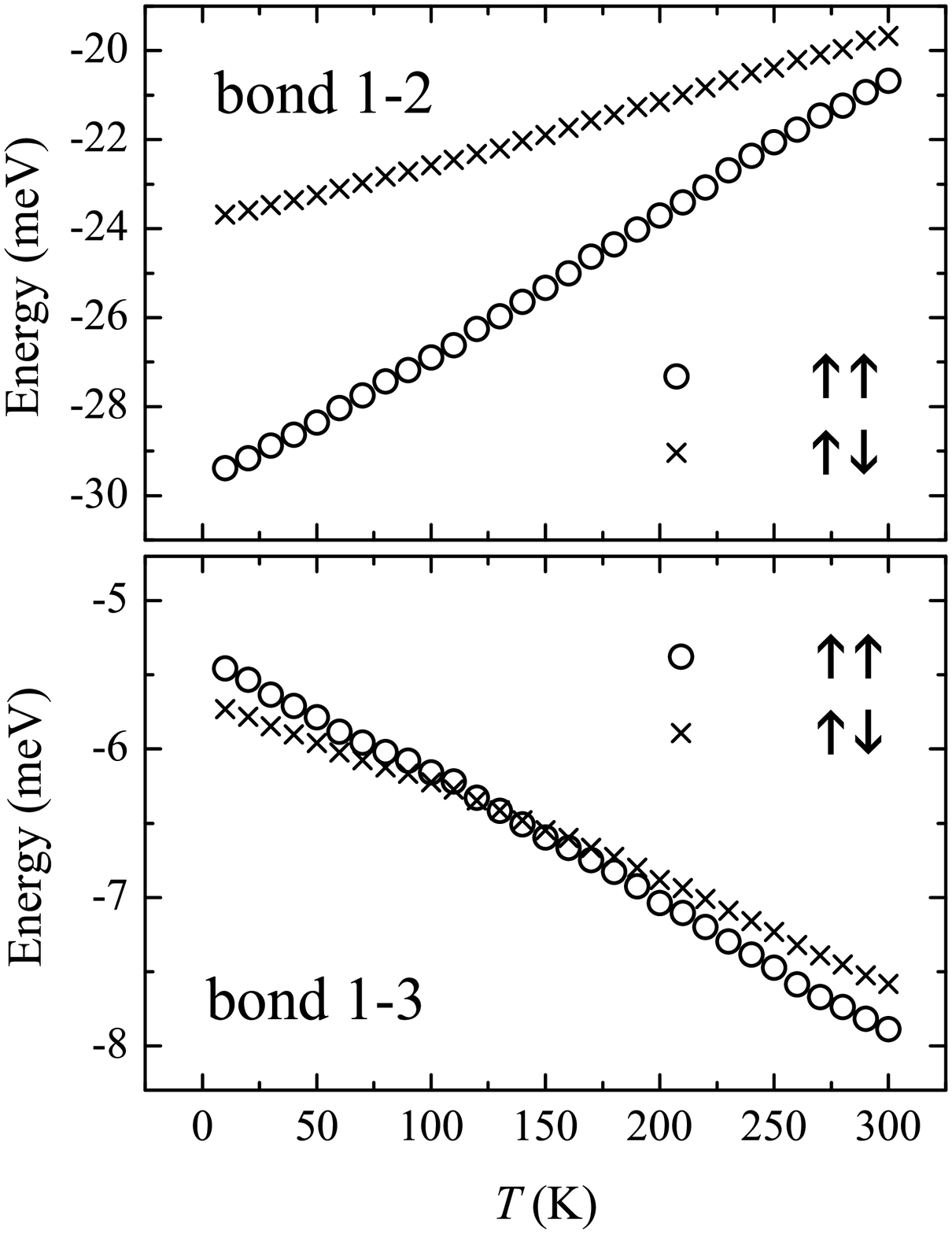}}
\resizebox{6.5cm}{!}{\includegraphics{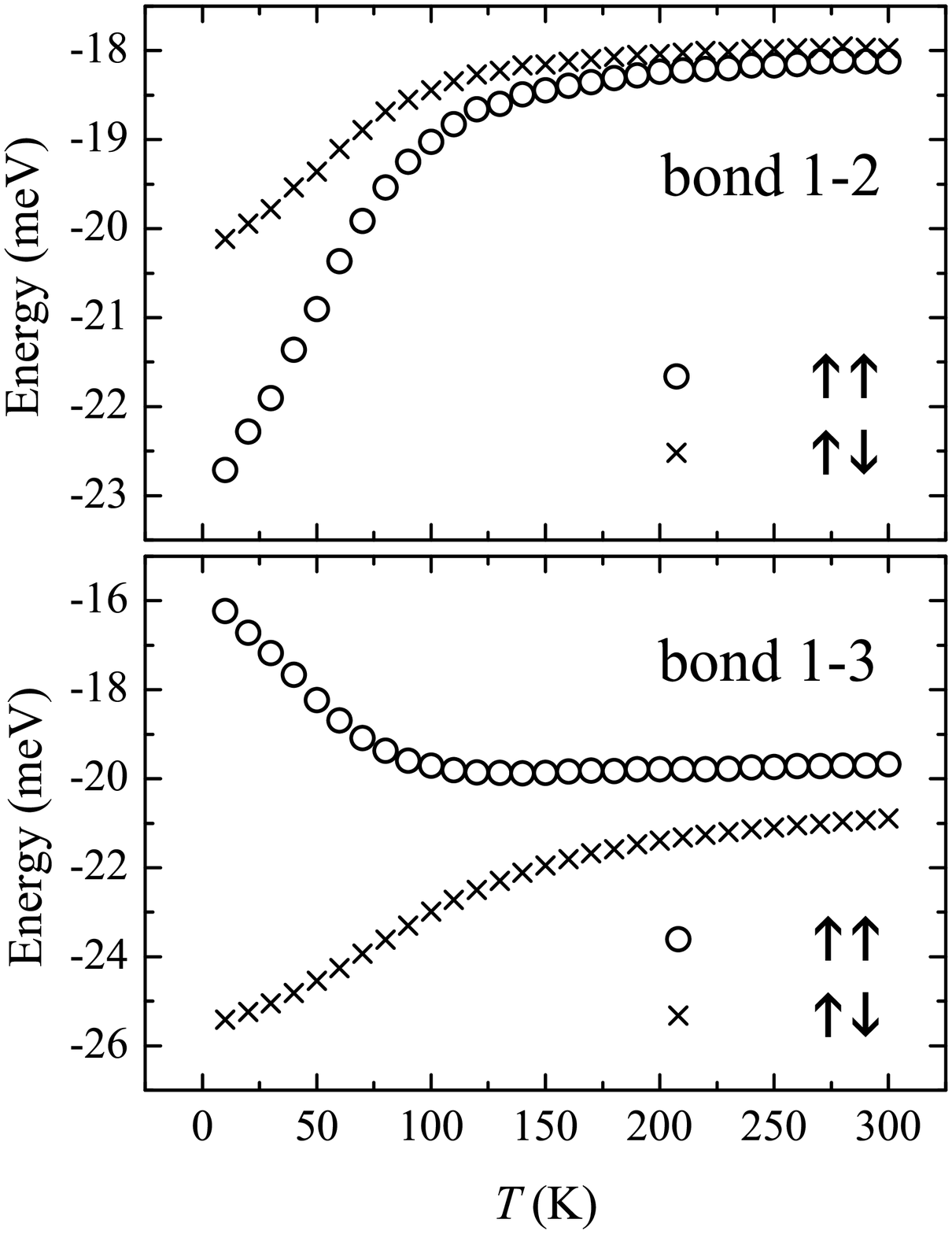}}
\caption{\label{fig.Taverage}
Thermal averages of pair interactions ${\cal T}_{ij}(T)$
for the ferromagnetic ($\uparrow \uparrow$) and
antiferromagnetic ($\uparrow \downarrow$) configurations of spins
in the bonds 1-2 and 1-3
calculated for the low-temperature structure of YTiO$_3$ (left)
and LaTiO$_3$ (right). The atomic positions are explained
in Fig.~\protect\ref{fig.structure}.
}
\end{figure}

  In this section we consider the effect of thermal
fluctuations
of the orbital degrees of freedom
on interatomic magnetic interactions between the spins.
For these purposes, it is convenient to work in the
local coordinate frame corresponding to the diagonal
representation of the
CF Hamiltonian $ \hat{h}^{\rm CF}_i $ at each atomic site.
Then, we assume that for each projection of spin,
the three-component occupied $t_{2g}$-orbital at the site $i$
can be presented in the form of the real vector
\begin{equation}
{\bi v}_i = \left(
\begin{array}{l}
\cos \theta_i \\
\sin \theta_i \cos \phi_i \\
\sin \theta_i \sin \phi_i
\end{array}
\right),
\label{eqn:rorbital}
\end{equation}
in the basis of three CF orbitals
listed in table \ref{tab:CF}, where
$0 \leq \theta_i \leq \pi/2$ and $0 \leq \phi_i \leq \pi$ due to the
invariance of both crystal field and pair interactions (\ref{eqn:egain})
with respect to the inversion ${\bi v}_i \rightarrow -$${\bi v}_i$.
In these notations, the point $\theta_i$$=$$\phi_i$$=$$0$
corresponds to the lowest
CF orbital. This situation was already
considered in Sec. \ref{sec:CF} in the limit of
large CF splitting. However, since the
CF splitting is finite, other orbital configurations can
contribute
to the thermodynamic averages of physical quantities
at elevated temperatures.
Thus, similar to Sec. \ref{sec:CF}, we assume that
the CF splitting is much larger that the
energy gain (\ref{eqn:egain}) caused by the virtual hoppings,
but can become comparable with the energies of thermal fluctuations,
and consider the finite-temperature extension of the
CF theory. Then, the thermal average of (\ref{eqn:egain})
is given by
\begin{equation}
\fl
{\cal T}_{ij}(T) = \int d\Omega_i
{\cal W}(\theta_i,\phi_i)
\int d\Omega_j
{\cal W}(\theta_j,\phi_j)
{\cal T}(\theta_i,\phi_i,\theta_j,\phi_j),
\label{eqn:Taverage}
\end{equation}
where
$$
\int d\Omega_i = \frac{1}{2\pi} \int_0^{2\pi} d\phi_i \int_0^{\pi/2} \sin\theta_i d\theta_i,
$$
$$
{\cal W}(\theta_i,\phi_i) =
\frac{1}{Z_i}
\exp \left\{-\frac{({\bi v}_i,\hat{h}^{\rm CF}_i {\bi v}_i)}{k_BT}\right\},
$$
$$
Z_i = \int d\Omega_i \exp \left\{-\frac{({\bi v}_i,\hat{h}^{\rm CF}_i {\bi v}_i)}{k_BT}\right\},
$$
and ${\cal T}(\theta_i,\phi_i,\theta_j,\phi_j)$ is the pair interaction (\ref{eqn:egain})
constructed
from the occupied $t_{2g}$-orbitals of the form (\ref{eqn:rorbital})
for either ferromagnetic ($\uparrow$$\uparrow$) or antiferromagnetic ($\uparrow$$\downarrow$)
configurations of spins
(in the following denoted as ${\cal T}^{\uparrow \uparrow}_{ij}$ and
${\cal T}^{\uparrow \downarrow}_{ij}$, respectively).
The numerical integration in (\ref{eqn:Taverage}) was performed
by using the the Metropolis algorithm \cite{Metropolis,Jorgensen}.
An example of the temperature dependencies of ${\cal T}^{\uparrow \uparrow}_{ij}(T)$ and
${\cal T}^{\uparrow \downarrow}_{ij}(T)$ for YTiO$_3$ and LaTiO$_3$
is shown in Fig.~\ref{fig.Taverage}.\footnote{
The behavior of GdTiO$_3$ and SmTiO$_3$ is rather similar to that
of YTiO$_3$ and not discussed here.
}
Then, using ${\cal T}_{ij}^{\uparrow \uparrow}(T)$ and
${\cal T}_{ij}^{\uparrow \downarrow}(T)$,
one can evaluate the temperature dependence of
interatomic interactions between the spins,
$2 {J}_{ij}(T)$$=$$
{\cal T}_{ij}^{\uparrow \uparrow}(T)$$-$$
{\cal T}_{ij}^{\uparrow \downarrow}(T)$, averaged
over all orbital configurations,
and self-consistently solve the equation for the magnetic
transition temperature in RPA (\ref{sec:appendixA}), where the value
of $T_{\rm C,N}$ is used as the
argument of $J_{ij}(T)$, and the procedure is repeated until
reaching the self-consistency with respect to $T_{\rm C,N}$.
Then, we obtain the following values of the Curie temperature
$T_{\rm C}$$=$ 60, 68, and 37 K for YTiO$_3$ (RT), GdTiO$_3$,
and SmTiO$_3$, respectively. YTiO$_3$ (LT) and LaTiO$_3$
are expected to develop the A-type AFM order with the
N\'eel temperature $T_{\rm N}$$=$ 64 and 52 K, respectively.
Thus, the thermal fluctuation systematically decreases the
values of the magnetic transition temperature (by 10-44 \%).
As expected, the largest change is observed in LaTiO$_3$,
which has the smallest CF-splitting, and in YTiO$_3$, due to the
quasi-two-dimensional character of interatomic interactions.
Nevertheless, SmTiO$_3$ remains FM and LaTiO$_3$ -- A-type
AFM, contrary to the experimental data.

\section{\label{sec:MultiSlater}Multi-Determinant Approach
for Superexchange Interactions}

  By summarizing results of the previous sections, we note the
following shortcomings of the one-electron approach:
\begin{itemize}
\item[$\bullet$]
In the case of LaTiO$_3$ and SmTiO$_3$, it fails to predict the correct G-type AFM ground state.
Although small G-type AFM component (along the ${\bf a}$-axis)
is expected in the theoretical magnetic ground state of
SmTiO$_3$ after including the relativistic SO interaction,
that of
LaTiO$_3$ does not involve the G-type AFM arrangement.
\item[$\bullet$]
Even for the FM compounds YTiO$_3$ and GdTiO$_3$, the magnetic transition
temperature is typically overestimated by factor two. To certain extent, it can
be reduced by thermal fluctuations of the orbital degrees of freedom.
On the other hand,
the relativistic SO interaction acts in the opposite direction and additionally
increases $T_{\rm C}$ (at least on the level of mean-field approximation).
\end{itemize}
In this section, we investigate whether these problems can be resolved
by considering the many-electron effects.

  As was pointed out in the beginning of Sec.~\ref{sec:SingleSlater},
the main drawback of all previous considerations was that,
although the expression (\ref{eqn:egain})
for the energy gain
takes into account the
many-electron effects in the intermediate configurations $t_{2g}^2$,
which can be reached in the process of virtual hoppings,
it was combined with the
one-electron approximation for the ground-state wavefunction
$G$ in the atomic limit.
The purpose of this section is to go beyond this
one-electron approximation and to clarify the role played by the
many-electron effects in the problem of SE interactions.
Namely, for each bond
$i$-$j$, we construct the
complete
basis of two-electron Slater determinants of the form,
\begin{equation}
|S(1,2) \rangle = \frac{1}{\sqrt{2}}
\left[ \psi_i(1) \psi_j(2) -
\psi_i(2) \psi_j(1) \right],
\label{eqn:SlaterBasis}
\end{equation}
where $\psi_i$ denotes the one-electron spin-orbital
residing
at the site $i$. In practice,
these orbitals were obtained from the diagonalization
of the CF Hamiltonian $\hat{h}^{\rm CF}_i$ (table~\ref{tab:CF}).
In total, there are $6$ such spin-orbitals and $36$
Slater determinants of the form (\ref{eqn:SlaterBasis}).
Then, for each bond, we calculate matrix elements
of the pair interactions
\begin{equation}
\fl
{\cal T}_{ij}^{SS'} = - \left\langle S
\left| \hat{t}_{ij} \left( \sum_M \frac{{\hat{\mathscr{P}}}_j|j M
\rangle \langle j M|
{\hat{\mathscr{P}}}_j}{E_{jM}} \right) \hat{t}_{ji} +
(i \leftrightarrow j) \right| S' \right\rangle
\label{eqn:egainManyS}
\end{equation}
in the basis of these Slater determinants,
and combine them with the matrix elements of one-electron
operators of the crystal field $\hat{h}^{\rm CF}_i + \hat{h}^{\rm CF}_j$
and (optionally) the SO interaction
$\hat{h}^{\rm SO}_i + \hat{h}^{\rm SO}_j$.
The corresponding $36$$\times$$36$ Hamiltonian matrix is denoted as
$\hat{H}_{ij}$$=$$\| H_{ij}^{SS'} \|$.
Then, the spectrum of two-electron states in each
bond is obtained after the diagonalization of $\hat{H}_{ij}$.
Thus, the basic idea (and approximation) behind this treatment is that the
entire lattice is divided into ``molecules'', and the electronic
structure of each ``molecule'' can be considered independently
from other ``molecules''. We will mainly focus on the origin of
AFM correlations in LaTiO$_3$.

\begin{figure}[h!]
\centering \noindent
\resizebox{5cm}{!}{\includegraphics{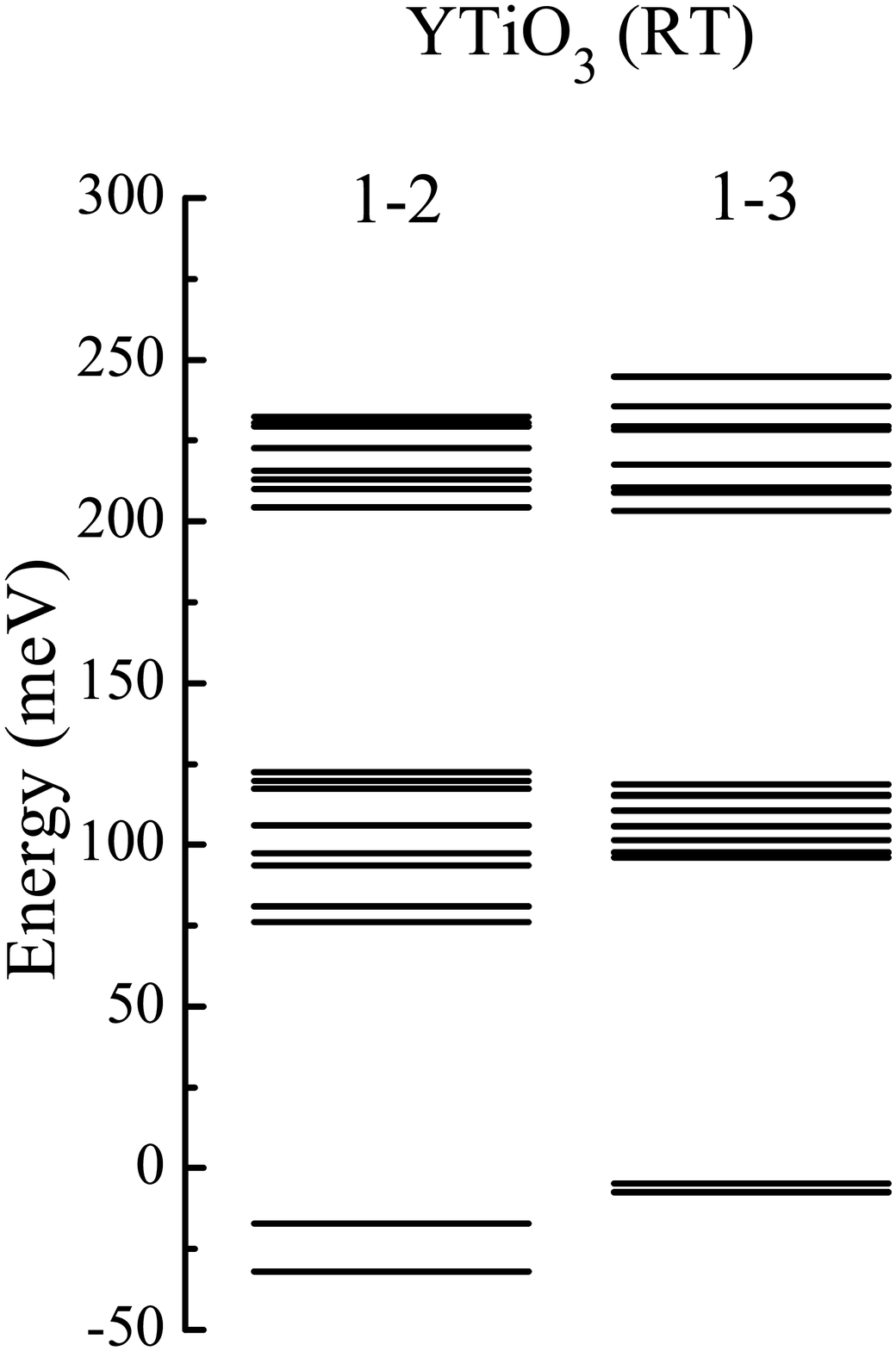}}
\resizebox{5cm}{!}{\includegraphics{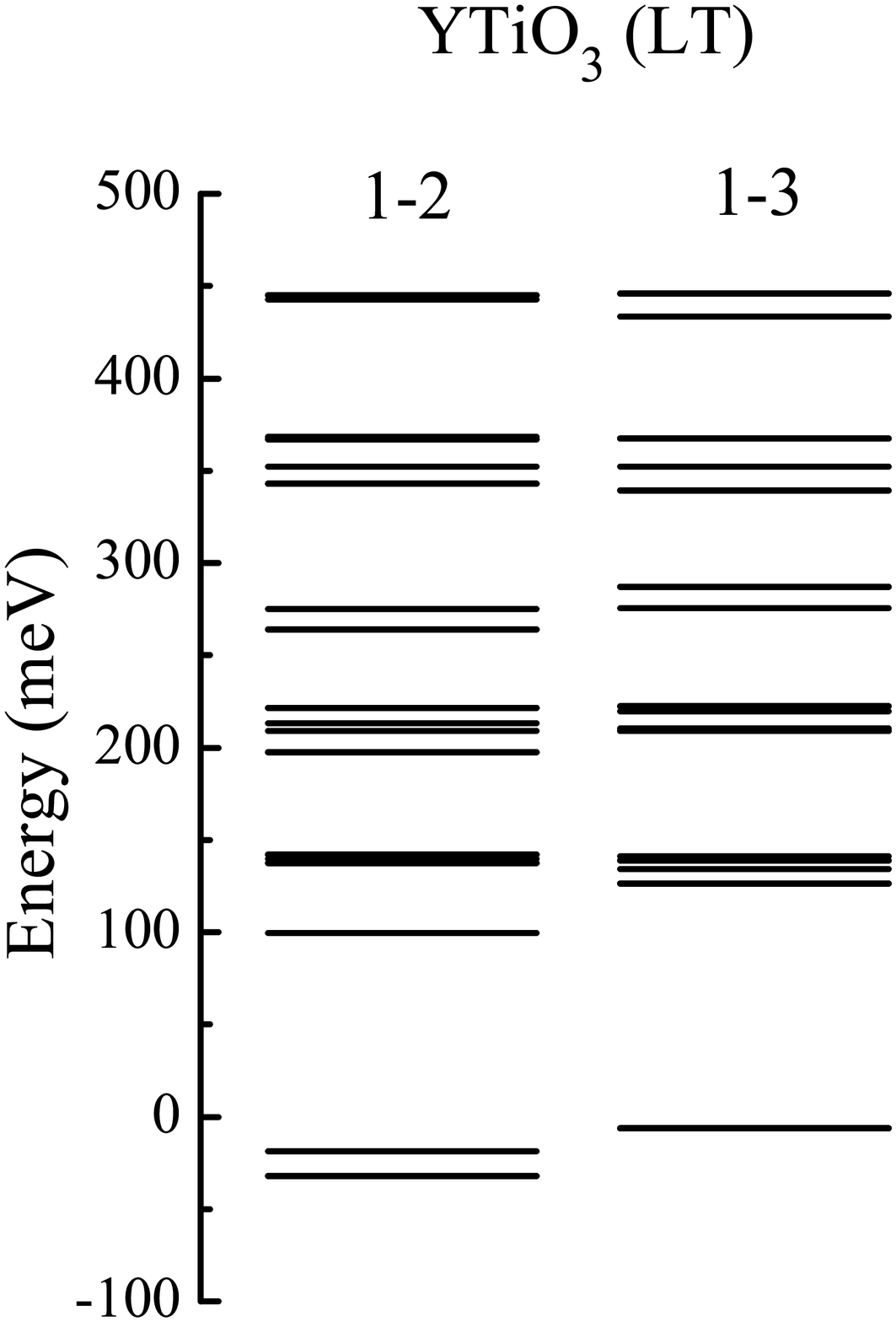}} \\
\resizebox{5cm}{!}{\includegraphics{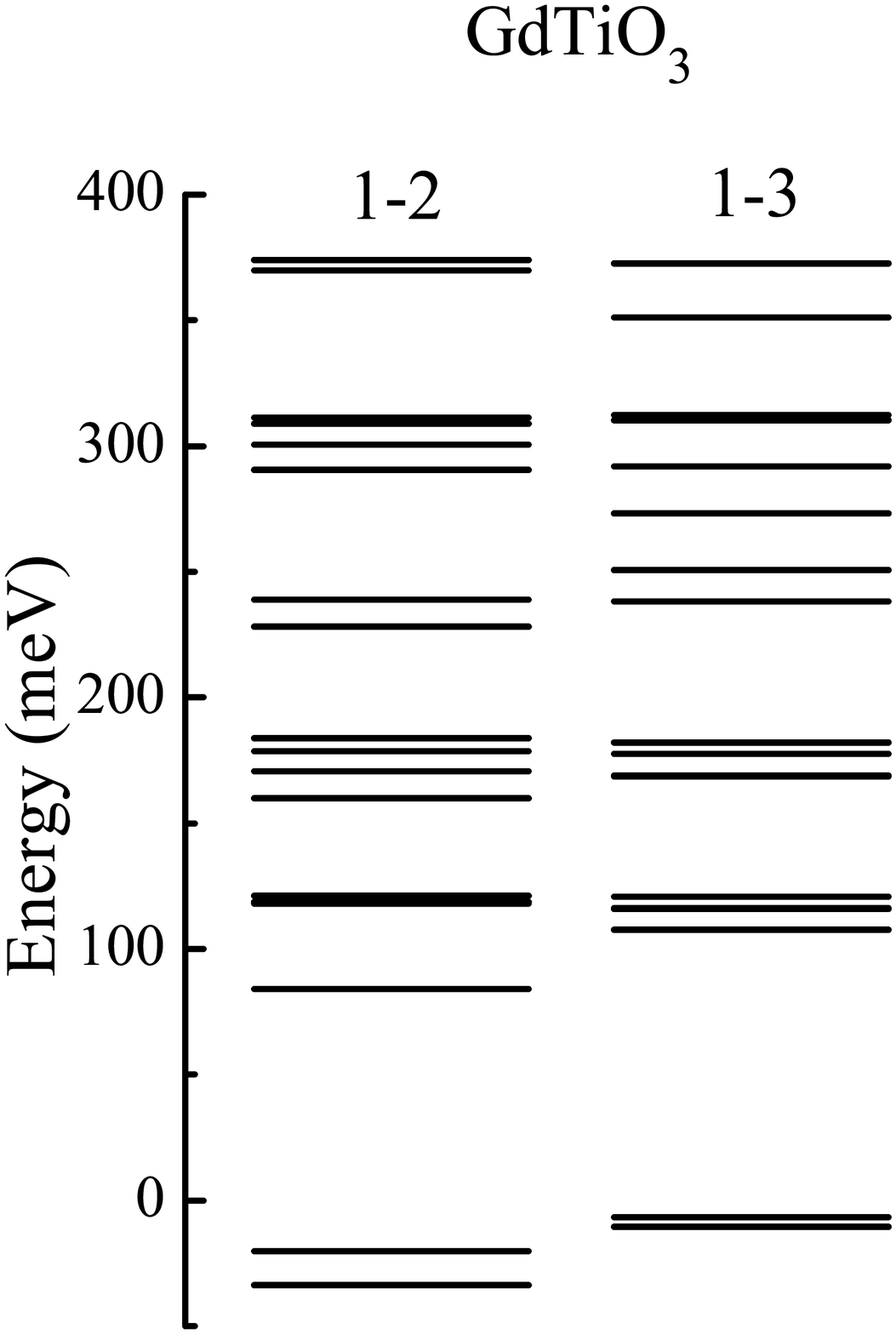}}
\resizebox{5cm}{!}{\includegraphics{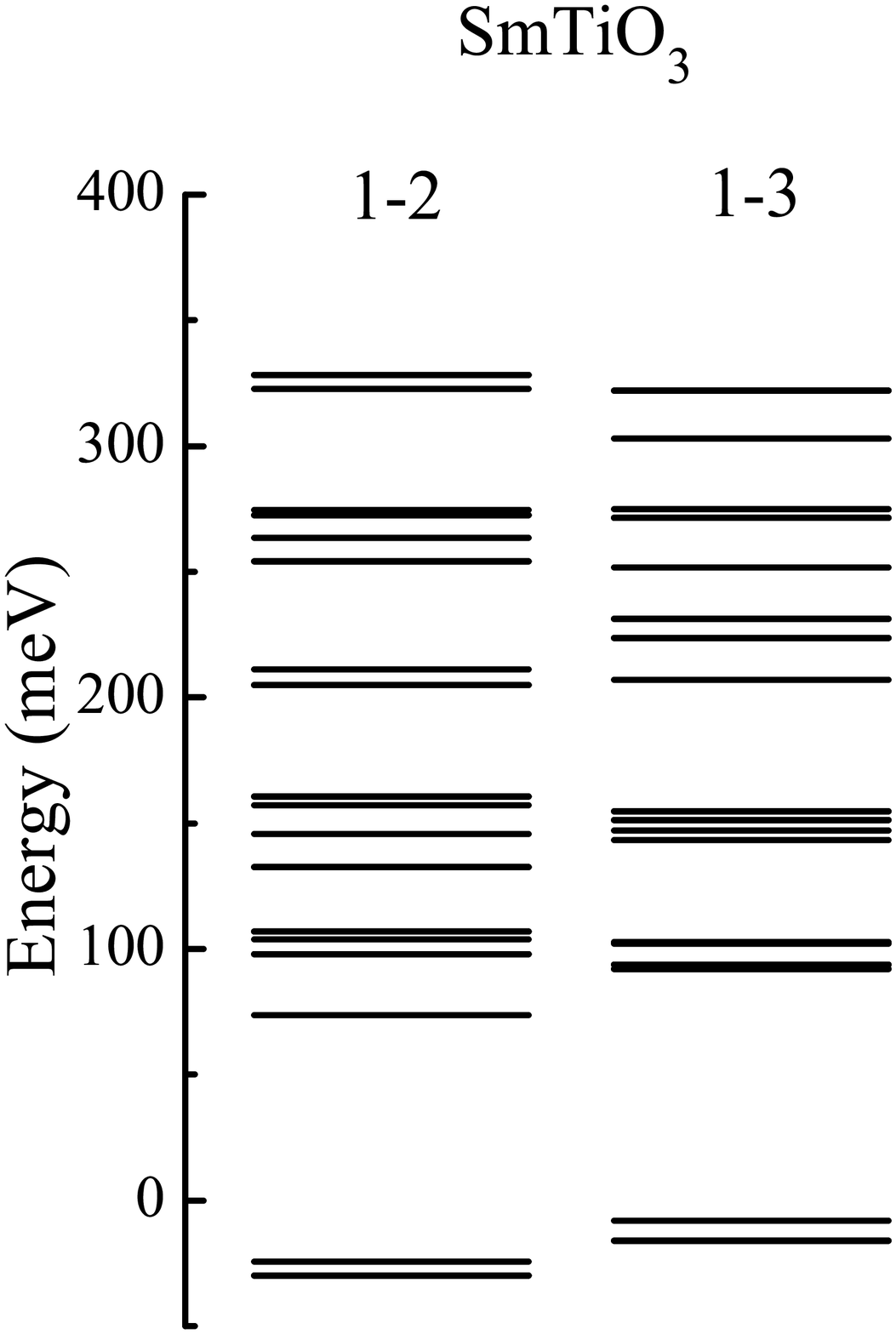}}
\resizebox{5cm}{!}{\includegraphics{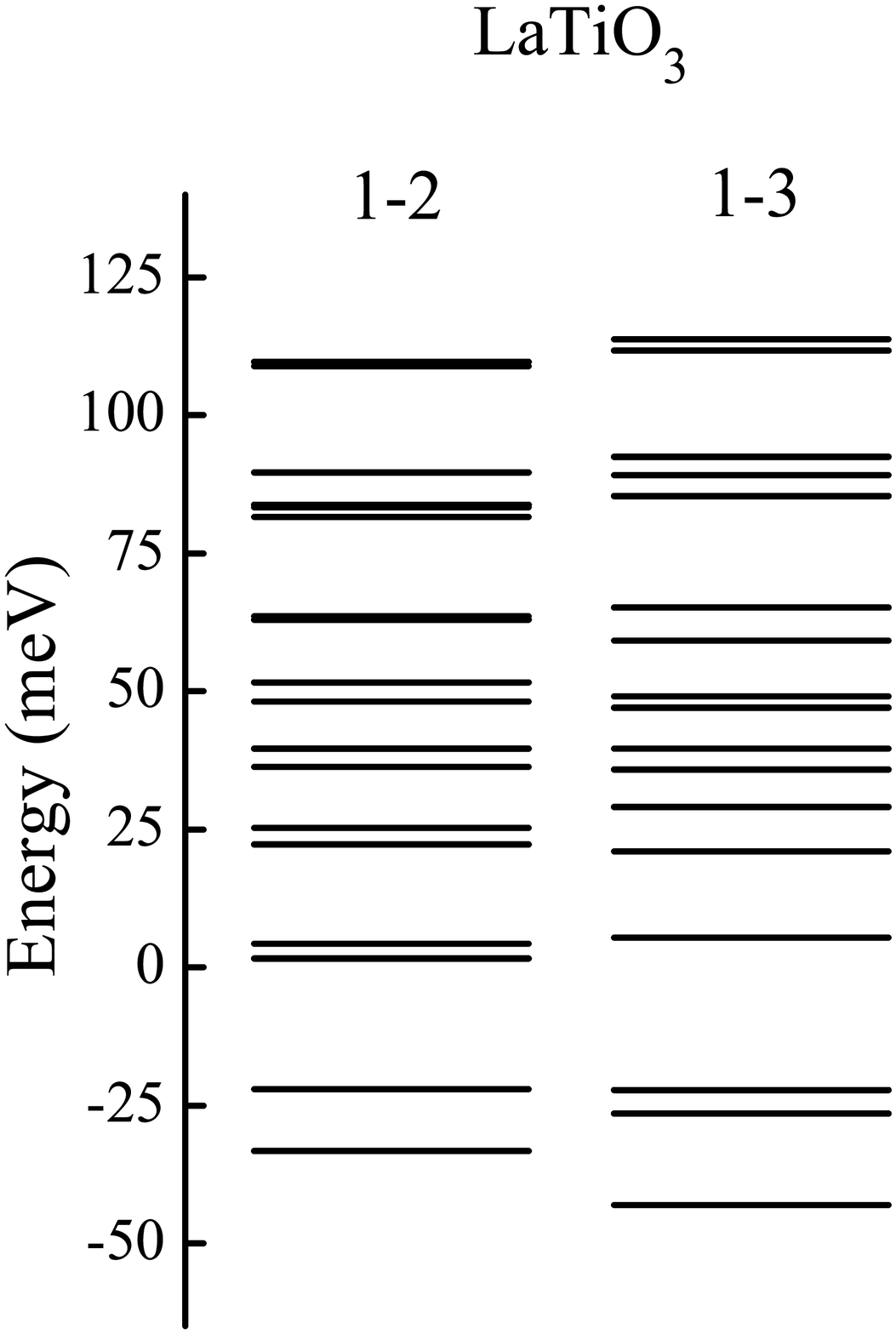}}
\caption{\label{fig.manySlater}
Spectrum of two-electron states obtained after the
diagonalization of pair and crystal-field interactions
in the complete basis of Slater determinants separately for the
bonds 1-2 and 1-3 (the atomic positions are explained in Fig.~\ref{fig.structure}).
}
\end{figure}

  First, let us discuss results without relativistic SO interaction.
The spectrum of two-electron states, obtained after the diagonalization
of $\hat{H}_{ij}$ for the bonds $1$-$2$ and $1$-$3$
is shown in Fig. \ref{fig.manySlater}.
There is a drastic difference of LaTiO$_3$ from other compounds.
In YTiO$_3$, SmTiO$_3$, and GdTiO$_3$, four low-energy levels,
including one spin-singlet and three degenerate spin-triplet states,
are clearly separated from the next states by an energy gap of
at least 100 meV, corresponding to the value of the
CF splitting (table \ref{tab:CF}). However, in LaTiO$_3$, the
singlet-triplet splitting in the low-energy part of the spectrum
is at least comparable with energy gap,
separating these levels from the next two-electron states.
For example, in the bond $1$-$3$, the singlet-triplet splitting is about
20 meV. However, the next level is located only 4 meV higher than the
triplet state. Thus, the crystal-field
theory, although applicable for YTiO$_3$
and high-temperature structures of SmTiO$_3$ and GdTiO$_3$,
definitely breaks down in the case of LaTiO$_3$, where the splitting of
the
two-electron levels caused by the SE effects
in the low-energy part of the spectrum is at least comparable
with the CF splitting.

  The parameter of the isotropic
exchange coupling in the bond $i$-$j$ can be expressed through the
energies of the low-lying spin-singlet ($E_S$)
and spin-triplet ($E_T$) states,
$$
J_{ij} = \frac{1}{4} \left( E_S - E_T \right),
$$
where the prefactor $1/4$ stands for $S^2$,
which according to (\ref{eqn:spinmodel2}) is already
included to the definition of $J_{ij}$.
The obtained parameters are listed in table \ref{tab:multiSJ} together with
the
values of the magnetic transition temperature $T_{\rm C,N}$ estimated
in the random phase approximation (\ref{sec:appendixA}).
\begin{table}[h!]
\caption{Parameters of isotropic exchange coupling
($J_{ij}$, measured in meV) derived from the
singlet-triplet splitting of the low-energy levels,
which were obtained
from the diagonalization of pair
and crystal-field interactions in the complete basis of
two-electron Slater determinants separately for each bond
of the system;
corresponding magnetic transition temperatures ($T_{\rm C,N}$, measured in K) obtained in the
random phase approximation; and the type of the magnetic ground state.
Depending on the magnetic ground state, the notations $T_{\rm C}$ and $T_{\rm N}$
stand for the Curie and N\'eel temperature, respectively.}
\label{tab:multiSJ}
\begin{indented}
\item[]\begin{tabular}{lcccccc}
\br
  compound     & $J_{12}$ & $J_{13}$ &  $J_{23}$ & $J_{23'}$ &  $T_{\rm C,N}$ & GS \\
\mr
YTiO$_3$ (RT)  &   $3.68$ & $\phantom{-}0.64$           & $0.03$ & $-0.16$ & 104    &  F \\
YTiO$_3$ (LT)  &   $3.34$ & $\phantom{-}0\phantom{.00}$ & $0.03$ & $-0.16$ & \phantom{1}79  & A \\
GdTiO$_3$      &   $3.39$ & $\phantom{-}0.98$           & $0.07$ & $-0.17$ & 116            & F \\
SmTiO$_3$      &   $1.43$ & $\phantom{-}2.00$           & $0.03$ & $\phantom{-}0.07$ & \phantom{1}87  & F \\
LaTiO$_3$      &   $2.82$ &           $-4.17$           & $0.30$ & $\phantom{-}0.14$ & 138            &  A \\
\br
\end{tabular}
\end{indented}
\end{table}
By comparing them with the values obtained in the one-electron
approximation (table \ref{tab:CFJ}), one can clearly see that the
two-electron effects tend to additionally stabilize the
FM interactions. Generally,
the values of all FM interaction $J_{ij}$
increase in the two-electron approach, while the magnitude
of the AFM interactions $J_{13}$ in LaTiO$_3$
and in the LT phase of YTiO$_3$
decreases. Particularly, $J_{13}$ nearly vanishes in the case of
YTiO$_3$ (LT), resulting in the two-dimensional character of NN
interactions. However, due to the next NN interactions between
the planes, $T_{\rm N}$ remains finite. Nevertheless,
only in YTiO$_3$ (LT)
the
two-electron effects tend to
decrease the magnetic transition temperature.
This behavior is solely related to the quasi-two-dimensional
character of interatomic magnetic interactions.

\begin{figure}[h!]
\centering \noindent
\resizebox{5cm}{!}{\includegraphics{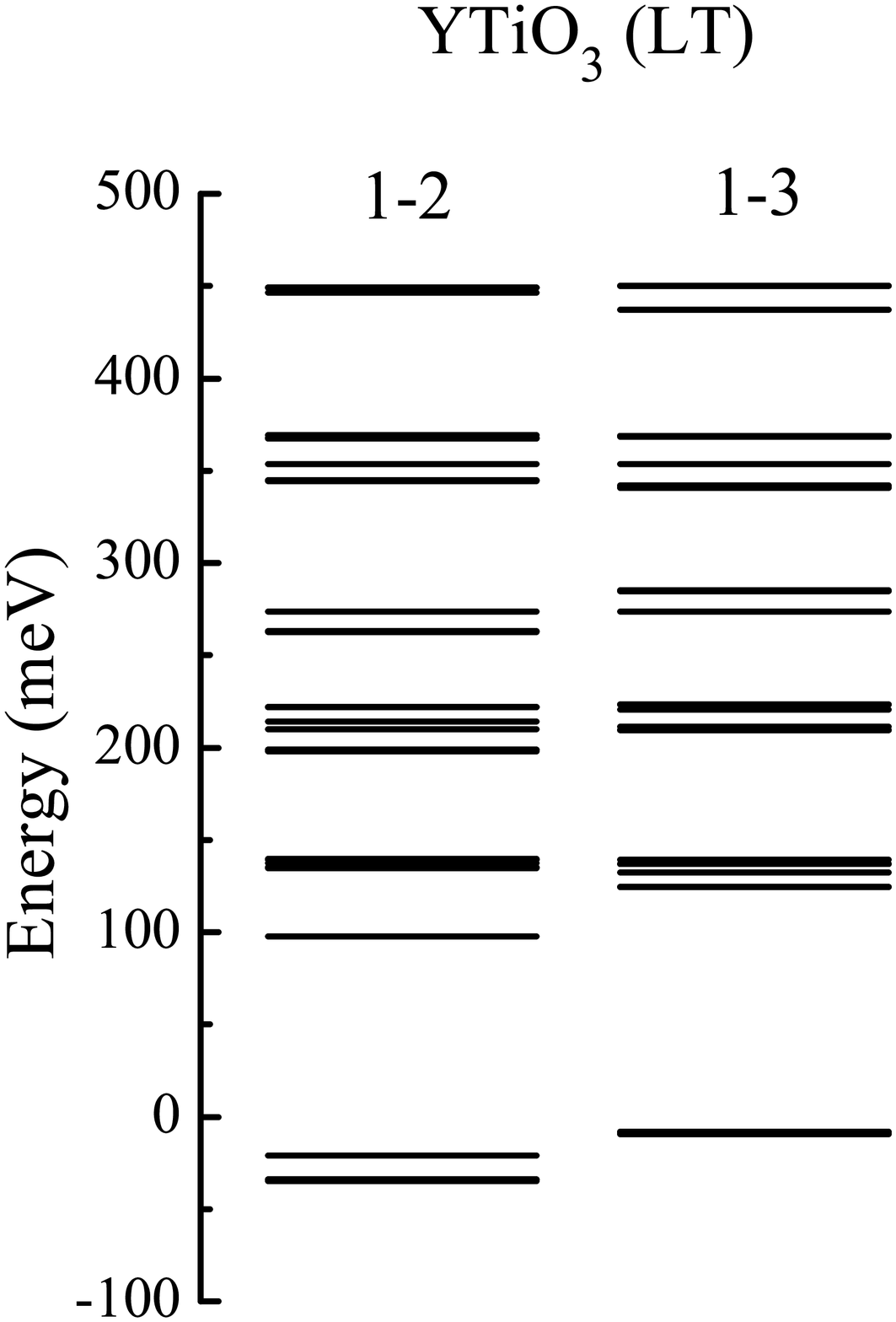}}
\resizebox{5cm}{!}{\includegraphics{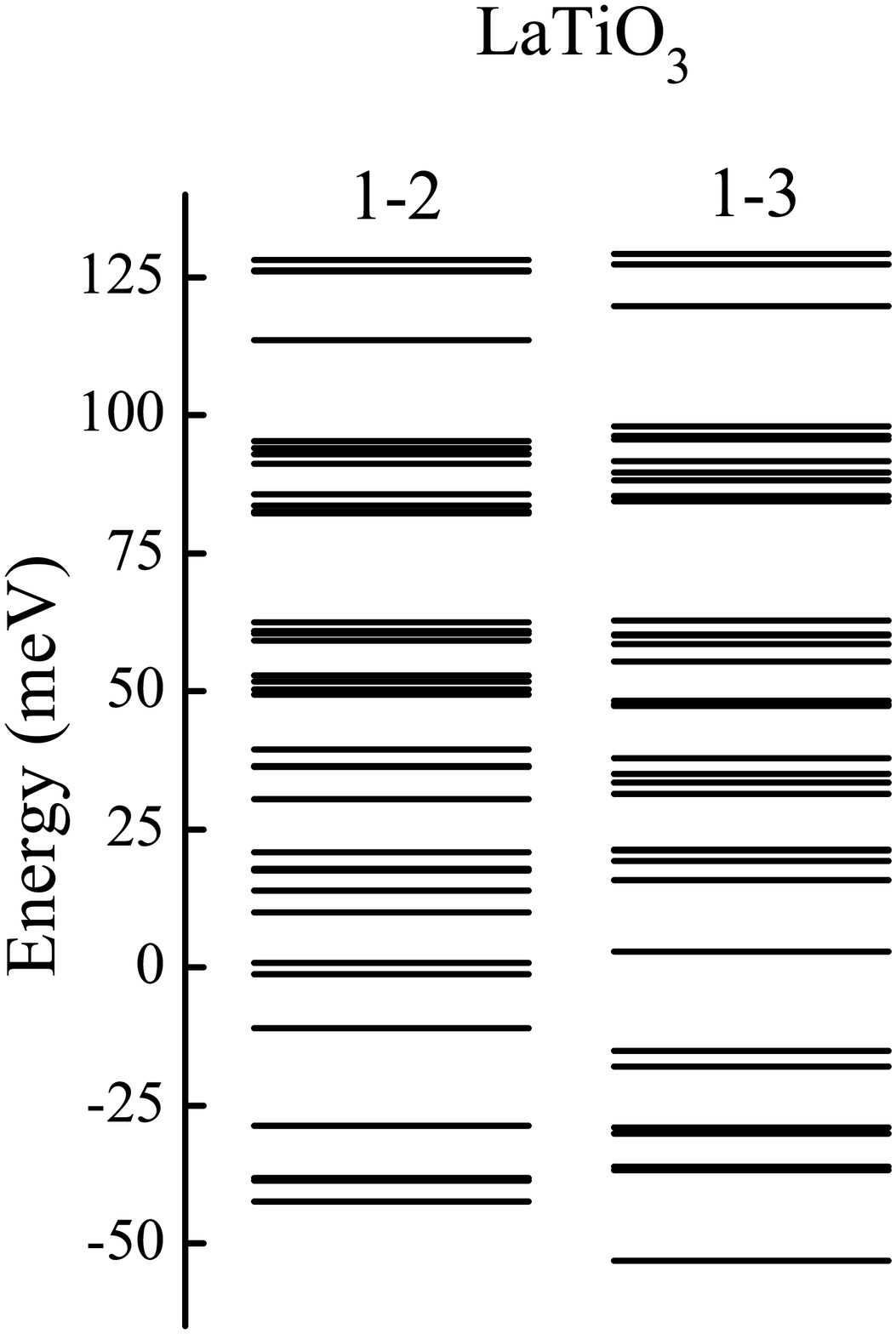}}
\caption{\label{fig.manySlaterR}
Spectrum of two-electron states obtained for the
low-temperature structure YTiO$_3$ (left)
and LaTiO$_3$ (right) after the
diagonalization of pair, crystal-field \textit{and}
relativistic spin-orbit interactions
in the complete basis of Slater determinants, separately for the
bonds 1-2 and 1-3 (the atomic positions are explained in Fig.~\ref{fig.structure}).
}
\end{figure}

  Thus, it seems unlikely that the two-electron effects
alone will stabilize the G-type AFM ground state in the case of LaTiO$_3$.
Therefore, we investigate the last possibility
related to the relativistic SO interaction.
For the most of the considered systems, the SO interaction is
small in comparison with the CF splitting and does not significantly change the
distribution of the two-electron states. The typical
example for the LT
structure of YTiO$_3$ is shown in Fig.~\ref{fig.manySlaterR}, which is practically identical to the
spectrum without the SO
interaction (Fig.~\ref{fig.manySlater}).
For the FM systems, the splitting of the low-lying spin-triplet states by the SO
interaction is very small. Nevertheless, below we will see that this splitting
may have rather interesting consequences on the magnetic properties.
Again, one clear exception is LaTiO$_3$, where the strength of the
SO interaction ($\xi = 21$ meV) is comparable with the CF splitting \emph{and} the
energy gain caused by the virtual hoppings,
and only in LaTiO$_3$
the spectrum of two-electron states
changes significantly when the relativistic SO interaction is taken into account.
In order to understand the character of magnetic interactions,
we calculate the parameters of interatomic spin correlations in the ground state
$$
\langle \hat{\sigma}_i^a \hat{\sigma}_j^b \rangle =
\langle \Psi | \hat{\sigma}_i^a \hat{\sigma}_j^b | \Psi \rangle,
$$
where $\Psi$ is the two-electron wavefunction for the bond $i$-$j$, corresponding to the
lowest eigenvalue of $\hat{H}_{ij}$.
These parameters are listed in table \ref{tab:Scorrelation}.
\begin{table}[h!]
\caption{Parameters of interatomic spin correlations in the ground state
calculated separately for each bond of the system after the diagonalization
of the two-electron Hamiltonian including pair, crystal field, and
relativistic spin-orbit interactions. The atomic positions
are explained in Fig.~\protect\ref{fig.structure}.}
\label{tab:Scorrelation}
\begin{indented}
\item[]\begin{tabular}{lcc}
\br
  compound     & $\| \langle \hat{\sigma}_1^a \hat{\sigma}_2^b \rangle \|$ &
$\| \langle \hat{\sigma}_1^a \hat{\sigma}_3^b \rangle \|$         \\
\mr
YTiO$_3$ (RT)  &
$
\left(
\begin{array}{ccc}
  \phantom{-}0.03 & -0.91 & -0.35 \\
  -0.92 & \phantom{-}0.08 & -0.29 \\
  -0.31 & -0.35 & \phantom{-}0.83 \\
\end{array}
\right)
$
&
$
\left(
\begin{array}{ccc}
  -0.40 & \phantom{-}0.83 &-0.29 \\
  \phantom{-}0.83 & \phantom{-}0.21 &-0.44 \\
  \phantom{-}0.29 & \phantom{-}0.44 & \phantom{-}0.77 \\
\end{array}
\right)
$
\\
\mr
YTiO$_3$ (LT)  &
$
\left(
\begin{array}{ccc}
  \phantom{-}0.10 & -0.95 & -0.22 \\
  -0.89 & \phantom{-}0\phantom{.00} & -0.43 \\
  -0.42 & -0.23 & \phantom{-}0.85 \\
\end{array}
\right)
$
&
$
\left(
\begin{array}{ccc}
  -0.91 & \phantom{-}0.25 & -0.28 \\
  \phantom{-}0.25 & -0.14 & -0.93 \\
  \phantom{-}0.28 & \phantom{-}0.93 & -0.07 \\
\end{array}
\right)
$
\\
\mr
GdTiO$_3$   &
$
\left(
\begin{array}{ccc}
  \phantom{-}0.30 & -0.80 & -0.47 \\
  -0.91 & -0.19 & -0.29 \\
  -0.16 & -0.53 & \phantom{-}0.79 \\
\end{array}
\right)
$
&
$
\left(
\begin{array}{ccc}
  -0.41 & \phantom{-}0.86 & -0.21 \\
  \phantom{-}0.86 & \phantom{-}0.33 & -0.33 \\
  \phantom{-}0.21 & \phantom{-}0.33 & \phantom{-}0.88 \\
\end{array}
\right)
$
\\
\mr
SmTiO$_3$   &
$
\left(
\begin{array}{ccc}
  \phantom{-}0.13 & -0.51 & -0.79 \\
  -0.92 & -0.23 & -0.06 \\
  \phantom{-}0.12 & -0.78 & \phantom{-}0.48 \\
\end{array}
\right)
$
&
$
\left(
\begin{array}{ccc}
  -0.66 & \phantom{-}0.71 & -0.06 \\
  \phantom{-}0.71 & \phantom{-}0.65 & -0.14 \\
  \phantom{-}0.06 & \phantom{-}0.14 &  \phantom{-}0.95 \\
\end{array}
\right)
$
\\
\mr
LaTiO$_3$   &
$
\left(
\begin{array}{ccc}
  -0.23 & -0.67 & \phantom{-}0.42 \\
  -0.44 & \phantom{-}0.28 & \phantom{-}0.59 \\
  \phantom{-}0.71 & \phantom{-}0.08 & \phantom{-}0.27 \\
\end{array}
\right)
$
&
$
\left(
\begin{array}{ccc}
  -0.62 & -0.01 & \phantom{-}0\phantom{.00} \\
  -0.01 & -0.55 & \phantom{-}0.16 \\
  \phantom{-}0\phantom{.00} & -0.16 & -0.90 \\
\end{array}
\right)
$
\\
\br
\end{tabular}
\end{indented}
\end{table}
The diagonal matrix elements
($a$$=$$b$)
describe the longitudinal spin correlations. If
$ \langle \hat{\sigma}_i^a \hat{\sigma}_j^a \rangle > 0$,
the correlations are FM.
If
$ \langle \hat{\sigma}_i^a \hat{\sigma}_j^a \rangle < 0$,
they are AFM.
The off-diagonal elements ($a$$\ne$$b$) describe
the transverse correlations, which are related to the noncollinear
spin arrangement in the bond $i$-$j$. Generally,
the longitudinal correlations are strongly anisotropic. For example,
within one bond the correlations can easily become either FM or AFM, depending on the
direction in the orthorhombic lattice.
The transverse correlations, which are solely caused by the relativistic SO
interaction, can be also strong and comparable with the longitudinal ones.

  For the FM systems, this picture may substantially differ from the one
without the SO interaction. However,
it should be noted that
the details of the
ground state in the FM case are determined by the small splitting of the
low-lying spin-triplet levels by the SO interaction. For the most of the
systems this splitting is small and typically varies from few hundredths of meV
to one meV. For example,
for the FM bond 1-2 in YTiO$_3$ (LT),
the second and
third levels are split from the lowest one by
0.56 meV and 0.59 meV, respectively. Therefore, when the temperature
exceeds
$T \sim 0.6$ meV$/k_B \approx 7$ K, the proper picture for the magnetic
interactions
in this compound
would correspond to the (thermal) average
$\| \langle \hat{\sigma}_1^a \hat{\sigma}_2^b \rangle \|$
over the low-lying ``spin-triplet'' states. For example, for the 1-2 bond in
YTiO$_3$ (LT) this procedure yields the following tensor
$$
\| \langle \hat{\sigma}_1^a \hat{\sigma}_2^b \rangle \|_{\rm avr} =
\left(
\begin{array}{ccc}
  \phantom{-}0.33 & \phantom{-}0.01 & -0.03 \\
  -0.01 & \phantom{-}0.33 & -0.04 \\
  \phantom{-}0.03 & \phantom{-}0.04 & \phantom{-}0.32 \\
\end{array}
\right).
$$
Similar results can be obtained for other systems forming the FM bonds.
After the averaging, the longitudinal
spin correlations become isotropic, while all transversal correlations
are considerably smaller. This behavior is consistent with the
form of
the interatomic magnetic interactions tensor
$\hat{A}_{ij}$ obtained on the level of one-electron
approximation (table~\ref{tab:Atensor}).
Another effect, which can mix the two-electron states within the lowest manifold
of certain bond is related to the virtual hopping in neighboring bonds, which
involve one of the atomic sites of the original bond.
Note that the many-electron effects will mix the lowest
CF orbital at certain atomic site with other states. The symmetry
of these states as well as the magnitude of the mixing
will be generally different for different bonds.
Since each atom participate
in several bonds, this effect will lead to the addition
mixing of the low-lying (spin-triplet) states obtained from
the diagonalization of the two-electron
Hamiltonian separately for each bond of the system.
Nevertheless, the quantitative estimate of this effect requires a
more rigorous solution of the many-electron problem, which is
beyond the scopes of the present work.

  In the case of LaTiO$_3$, all longitudinal correlations in the bond 1-3
are antiferromagnetic, while the transversal correlations are small.
The correlations in the bond 1-2 are strongly anisotropic:
if $yy$- and $zz$-components favor the FM coupling,
the $xx$-correlations are
antiferromagnetic.
The splitting between the lowest and the next two two-electron levels
in the bond 1-2 is about 4 meV, which is considerably larger than
the splitting obtained
in other (more distorted) compounds. Therefore, the effect is expected to be
more robust against the thermal fluctuations
mixing the spin-triplet states.
Strong transversal correlations are also expected.
Thus, the $xx$-correlations appears to be antiferromagnetic
simultaneously in the
bonds 1-2 and 1-3, being consistent with the
G-type AFM structure. On the basis of this analysis,
we expect that if the G-type AFM order took place in LaTiO$_3$,
it would be more likely developed by the $x$- (${\bf a}$-) projections of the
magnetic moments in the orthorhombic coordinate frame.

\section{\label{sec:Conclusions} Summary and Conclusions}

  Starting from the
multiorbital
Hubbard Hamiltonian
for the $t_{2g}$-bands of orthorhombically distorted titanates
$R$TiO$_3$ ($R$$=$ Y, Gd, Sm, and La), where all the
parameters were derived from the first-principles electronic structure
calculations,
we considered different levels of approximations for the construction
of the spin-only superexchange model.
Namely, after considering the conventional crystal-field theory, where
all interatomic magnetic interactions are solely determined by
occupied $t_{2g}$-orbitals, that are split off by the
crystal distortion, we consecutively incorporate the effects of the
relativistic spin-orbit interaction, thermal fluctuations of the
orbital degrees of freedom, and many-electron effects related to the
virtual electron hoppings in the bonds.

  Even in the conventional CF theory,
the interatomic magnetic interactions appear to be extremely sensitive to
the details of
the orbital structure, and small change of the orbital structure,
caused by either
crystal distortions or thermal fluctuations, can have a profound
effect on the magnetic properties.

  Particularly, the use of the
room- and low-temperature structure for YTiO$_3$ provides rather different
sets of parameters of the spin Hamiltonian. The additional distortion
in the low-temperature structure 
tends to weaken the ferromagnetic interactions:
it substantially reduces the FM coupling $J_{12}$
in the orthorhombic ${\bf ab}$-plane
and makes the coupling along the ${\bf c}$-axis weakly AFM.
Similar tendency was reported for LaTiO$_3$, where
the use of
the low-temperature structure \cite{Cwik} could help to
stabilize the G-type AFM state \cite{MochizukiImada,Schmitz,Okatov}.
Apparently, some of the problems encountered in the present work,
such as
the incorrect
magnetic ground state in SmTiO$_3$ and the overestimation of the Curie
temperature in GdTiO$_3$, may be resolved
by using the low-temperature structural data for
these compounds, which are not available today.

  Moreover, the thermal fluctuations of the
of the orbital degrees of freedom near the CF configuration
can substantially reduce the value of the magnetic transition temperature
(up to 40\%).

  The relativistic spin-orbit interaction,
which is responsible for the appearance of
anisotropic and antisymmetric Dzyaloshinsky-Morita interactions,
leads to the noncollinear magnetic alignment. The role of the
relativistic effects increases in the direction
YTiO$_3$$\rightarrow$GdTiO$_3$$\rightarrow$SmTiO$_3$$\rightarrow$LaTiO$_3$,
when the crystal distortion decreases.

  The crystal-field theory, although applicable for YTiO$_3$ and 
high-temperature structures of GdTiO$_3$
and SmTiO$_3$, definitely breaks down in the case of LaTiO$_3$, which has
the smallest CF-splitting and where other factors, such as the relativist
SO interaction and many-electron effects in the bonds, start to play an
important role. Particularly, we found that the combination of the latter
two factors could explain the G-type AFM character of interatomic
correlations in LaTiO$_3$.

\ack
I wish to thank
Alexey Mozhegorov and Yukitoshi Motome for useful discussions.
This work is partly supported by Grant-in-Aid for Scientific
Research in Priority Area ``Anomalous Quantum Materials''
and Grant-in-Aid for Scientific
Research (C) No. 20540337
from the
Ministry of Education, Culture, Sport, Science and Technology of
Japan.

\appendix
\section{\label{sec:appendixA}Random-Phase Approximation for the
Magnetic Transition Temperature in Many-Atomic Case}

  In this appendix, we present the generalization of the
well-known expression for the magnetic transition (Curie or N\'eel)
temperature in the
random phase approximation (RPA):\footnote{also known as the Tyablikov approximation,
renormalized spin-wave theory, or the spherical Heisenberg model \cite{Nagaev,Tyablikov,Mattis}.
}
\begin{equation}
T_{\rm C,N} = \frac{S(S+1)}{3k_B S^2} \left( \sum_{\bf q}
\frac{1}{J({\bf Q}) - J({\bf q})}  \right)^{-1},
\label{eqn:TRPAsimple}
\end{equation}
where $J({\bf q})$ is the Fourier image of $J_{ij}$
and
${\bf Q}$ is the vector describing the magnetic structure in the ground state,
to the case of the complex lattices containing $n$ magnetic atoms in the primitive cell.

  First, we transform the parameters of the Heisenberg model to the local
coordinate frame:
$J_{ij} \rightarrow \widetilde{J}_{ij} = s_{ij} J_{ij}$, where $s_{ij}$$=$ $+$$1$ and $-$$1$
corresponds to the FM and AFM arrangement of spins in the bond $i$-$j$
in the magnetic ground state.
For a collinear magnetic configuration,
this procedure is equivalent to the shift of the origin of the
Brillouin zone in the right-hand side of (\ref{eqn:TRPAsimple}). Then, we construct
the dynamical matrix:
$$
\widehat{\mathfrak J}({\bf q}) = \| \widetilde{J}_\ell \delta_{\ell \ell'} -
\widetilde{J}_{\ell \ell'}({\bf q}) \|,
$$
where
$$
\widetilde{J}_{\ell \ell'}({\bf q}) = \frac{1}{\cal N} \sum_{i \in \ell} \sum_{j \in \ell'} \widetilde{J}_{ij}
e^{i({\bf q},{\bf R}_i-{\bf R}_j)}
$$
is the Fourier image of $\widetilde{J}_{ij}$ acting between
the atomic sublattices $\ell$ and $\ell'$, ${\cal N}$ is the number of the primitive cells,
$$
\widetilde{J}_\ell = \sum_{\ell' = 1}^n \widetilde{J}_{\ell \ell'}(0),
$$
and ${\bf R}_i$ is the radius-vector of the site $i$.
Then,
by diagonalizing $\widehat{\mathfrak J}({\bf q})$, one can find its
eigenvalues $\omega_\ell({\bf q})$ (the ``magnon energies'') and by repeating the RPA
arguments \cite{Tyablikov} derive the following expression for the
magnetic transition temperature:
$$
T_{\rm C,N} = \frac{nS(S+1)}{3k_B S^2} \left( \sum_{\bf q} \sum_{\ell = 1}^n 1/\omega_\ell({\bf q}) \right)^{-1}.
$$

\section{\label{sec:appendixB}Mean-Field Approximation for the
Magnetic Transition Temperature in the Case of the Noncollinear Spin Arrangement}

  The mean-field approximation for the relative magnetization $\bsigma_i (T)$,
which is the temperature average of $\hat{\bsigma}_i$, is formulated
in the following way. The mean field (or the molecular field), corresponding
to the spin Hamiltonian (\ref{eqn:spinmodel1}), is given by
$$
{\bi h}_i = \sum_j \hat{A}_{ij} \bsigma_j (T).
$$
Then, the temperature average of $\hat{\bsigma}_i$ in the molecular field
${\bi h}_i$ has the following form
\begin{equation}
\bsigma_i (T) = \frac{{\bi h}_i}{|{\bi h}_i|} B_S \left( \frac{|{\bi h}_i|}{k_BT} \right),
\label{eqn:SMeanField}
\end{equation}
where $B_S$ is the Brillouin function for the spin $S$$=$$1/2$~\cite{Mattis}.
The equation (\ref{eqn:SMeanField}) is solved self-consistently. The transition temperature
is defined as the minimal temperature for which $\bsigma_i (T) = 0$.


\section*{References}
\begin{thebibliography}{10}

\bibitem{Greedan}
Greedan J E 1985
{\it J. Less-Common Met.} {\bf 111} 335

\bibitem{Komarek07}
Komarek A C, Roth H, Cwik M, Stein W-D, Baier J,
Kriener M, Bour\'ee F, Lorenz T and Braden M 2007
{\it Phys. Rev.} B {\bf 75} 224402

\bibitem{Knafo}
Knafo W, Meingast C, Boris A V, Popovich P, Kovaleva N N, Yordanov P,
Maljuk A, Kremer R K, L\"ohneysen H v and Keimer B 2009
{\it Phys. Rev.} B {\bf 79} 054431

\bibitem{KhaliullinMaekawa}
Khaliullin G and Maekawa S 2000
{\it Phys. Rev. Lett.} {\bf 85} 3950

\bibitem{MochizukiImada}
Mochizuki M and Imada M 2004
{\it New J. Phys.} {\bf 6} 154

\bibitem{PavariniNJP}
Pavarini E, Yamasaki A, Nuss J and Andersen O K 2005
{\it New Journal of Physics} {\bf 7} 188

\bibitem{Schmitz}
Schmitz R, Entin-Wohlman O, Aharony A, Harris A B and
M\"{u}ller-Hartmann E 2005
{\it Phys. Rev.} B {\bf 71} 144412

\bibitem{review2008}
Solovyev I V 2008
{\it J. Phys.: Condens. Matter} {\bf 20} 293201

\bibitem{PRB04}
Solovyev I V 2004
{\it Phys. Rev.} B {\bf 69} 134403

\bibitem{PRB06b}
Solovyev I V 2006
{\it Phys. Rev.} B {\bf 74} 054412

\bibitem{PRB06a}
Solovyev I V 2006
{\it Phys. Rev.} B {\bf 73} 155117

\bibitem{Maclean}
Maclean D A, Ng H-N and Greedan J E 1979
{\it J. Solid State Chem.} {\bf 30} 35

\bibitem{Cwik}
Cwik M, Lorenz T, Baier J, M\"{u}ller R, Andr\'{e} G,
Bour\'{e}e F, Lichtenberg F, Freimuth A, Schmitz R,
M\"{u}ller-Hartmann E and Braden M 2003
{\it Phys. Rev.} B {\bf 68} 060401(R)

\bibitem{Kanamori}
Kanamori J 1963
{\it Prog. Theor. Phys.} {\bf 30} 275

\bibitem{PRL05}
Solovyev I V 2005
{\it Phys. Rev. Lett.} {\bf 95} 267205

\bibitem{PWA}
Anderson P W 1959
{\it Phys. Rev.} {\bf 115} 2

\bibitem{KugelKhomskii}
Kugel K I and Khomskii D I 1982
{\it Sov. Phys. Usp.} {\bf 25} 231

\bibitem{KO2}
Solovyev I V 2008
{\it New J. Phys.} {\bf 10} 013035

\bibitem{MerminWagner}
Mermin N D and Wagner H 1966
{\it Phys. Rev. Lett.} {\bf 17} 1133

\bibitem{Nagaev}
Nagaev E L 1988
\textit{Magnetics with Complex Exchange Interactions}
(Moscow: Nauka)

\bibitem{Ulrich2002}
Ulrich C, Khaliullin G, Reehuis M, Ivanov A, He H, Taguchi Y,
Tokura Y and Keimer B 2002
{\it Phys. Rev. Lett.} {\bf 89} 167202

\bibitem{Akimitsu}
Akimitsu J, Ichikawa H, Eguchi N, Miyano T,
Nishi M and Kakurai K 2001
{\it J. Phys. Soc. Jpn.} {\bf 70} 3475

\bibitem{SawadaTerakura}
Sawada H and Terakura K 1998
{\it Phys. Rev.} B {\bf 58} 6831

\bibitem{Metropolis}
Metropolis N, Rosenbluth A W, Rosenbluth M N,
Teller A H and Teller E 1953
{\it J. Chem. Phys.} {\bf 21} 1087

\bibitem{Jorgensen}
Jorgensen W L 2000
{\it Theor. Chem. Accounts} {\bf 103} 225

\bibitem{Okatov}
Okatov S, Poteryaev A and Lichtenstein A 2005
{\it Europhys. Lett.} {\bf 70} 499

\bibitem{Tyablikov}
Tyablikov S V 1975
\textit{Methods of Quantum Theory of Magnetism}
(Moscow: Nauka)

\bibitem{Mattis}
Mattis D C 2006
\textit{The Theory of Magnetism Made Simple}
(Singapore: World Scientific)



\end{thebibliography}
\end{document}